\documentclass[camera,hyphens]{jpaper}
\usepackage{amsmath,amsfonts}
\usepackage{graphicx}
\usepackage{textcomp}
\usepackage{fancyhdr}
\usepackage{graphicx}
\usepackage{textcomp}
\usepackage{xspace}
\usepackage{setspace}

\usepackage{comment}
\usepackage{listings}
\usepackage{caption}
\usepackage{algorithm}
\usepackage[noend]
{algpseudocode}
\usepackage{clipboard}
\renewcommand{\algorithmiccomment}[1]{\bgroup\hfill\fontsize{5.8}{6}\selectfont$\triangleright$ {\textcolor{blue}{#1}}\egroup}
\usepackage{booktabs} 
\usepackage{graphicx}
\usepackage{textcomp}
\usepackage{balance}
\usepackage{tikz}
\usepackage[rightcaption]{sidecap} 
\usepackage{xcolor, colortbl}
\usepackage[11pt]{moresize}
\usepackage{array,graphicx}

\usepackage{svg}
\usepackage{pifont}
\usepackage{threeparttable}
\usepackage{graphics}
\usepackage{multirow}
\usepackage{tabularx}
\usepackage{verbatim}
\usepackage{booktabs}
\usepackage{graphicx, caption, subcaption}

\usepackage{layout}
\usepackage[us,12hr]{datetime}
\usepackage{notoccite}
\usepackage{cite}
\usepackage{anyfontsize}
\usepackage{soul,color}
\usepackage{svg}
\usepackage{amsfonts}
\usepackage{fancyhdr}
\usepackage{datetime}
\usepackage[acronym]{glossaries}
\usepackage{sidecap}
\sidecaptionvpos{figure}{t} 
\usepackage[colorinlistoftodos]{todonotes}
\usepackage[hang,flushmargin]{footmisc} 
\usepackage[colorlinks,urlcolor=black,linkcolor=black,citecolor=black]{hyperref}

\usepackage[most]{tcolorbox}
\usepackage{array, booktabs,tabularx, longtable}
\definecolor{dollarbill}{rgb}{0.52, 0.99, 0.4}
\definecolor{dred}{rgb}{0.75, 0.00, 0.00}
\definecolor{dgreen}{rgb}{0.00, 0.5, 0.00}
\definecolor{ddgreen}{rgb}{0.00, 0.50, 0.00}
\definecolor{dpink}{rgb}{0.75, 0.0, 0.75}
\definecolor{dblack}{rgb}{0.00, 0.00, 0.00}
\definecolor{dblue}{rgb}{0.00, 0.00, 0.75}
\definecolor{gyell}{rgb}{0.5, 0.5, 0.0}
\definecolor{dbleudefrance}{rgb}{0.19, 0.55, 0.91}
\definecolor{darkgoldenrod}{rgb}{0.72, 0.53, 0.04}
\definecolor{NavyBlue}{rgb}{0.19, 0.55, 0.91}
\definecolor{byzantine}{rgb}{0.84, 0.2, 0.64}
\definecolor{amber}{rgb}{1.0, 0.75, 0.0}
\definecolor{canaryyellow}{rgb}{1.0, 0.94, 0.0}
\definecolor{amethyst}{rgb}{0.6, 0.4, 0.8}
\newcommand{\gs}[1]{{\color{black}{#1}}}
\newcommand{\gss}[1]{{\color{black}{#1}}}
\newcommand{\gsss}[1]{{\color{black}{#1}}}
\newcommand{\gssss}[1]{{\color{black}{#1}}}
\newcommand{\gsssss}[1]{{\color{black}{#1}}}
\newcommand{\gssssss}[1]{{\color{black}{#1}}}

\newcommand{\go}[1]{{\color{black}{#1}}}
\newcommand{\hc}[1]{{\color{black}{#1}}}
\newcommand{\gont}[1]{{\color{black}{#1}}}

\newcommand{\sr}[1]{{\color{black}{#1}}}

\DeclareMathOperator*{\argmax}{arg\,max}
\DeclareMathOperator*{\argmin}{arg\,min}
\newcommand{\cc}{\texttt{Causalcall}\xspace}
\newcommand{\gp}{{\texttt{Bonito\_CRF-sup}}\xspace}
\newcommand{\gpf}{{\texttt{Bonito\_CRF-fast}}\xspace}
\newcommand{\bon}{{\texttt{Bonito\_CTC}}\xspace}
\newcommand{\dor}{\texttt{Dorado-fast}\xspace}
\newcommand{\sac}{\texttt{SACall}\xspace}
\newcommand{\framework}{\texttt{RUBICON}\xspace}
\newcommand{\mech}{\texttt{RUBICALL}\xspace}
\newcommand{\mechmp}{\texttt{RUBICALL-MP}\xspace}
\newcommand{\mechfp}{\texttt{RUBICALL-FP}\xspace}
\newcommand{\nas}{\texttt{QABAS}\xspace}
\newcommand{\strim}{\texttt{SkipClip}\xspace}

\newcommand*\circled[1]{\tikz[baseline=(char.base)]{
            \node[shape=circle,draw,inner sep=0pt,fill=black, text=white] (char) {#1};}}

	\makeatletter
	\g@addto@macro{\normalsize}{%
	  \setlength{\abovedisplayskip}{1pt plus 1pt minus 1pt}
	  \setlength{\belowdisplayskip}{1pt plus 1pt minus 1pt}
	  \setlength{\abovedisplayshortskip}{0pt}
	  \setlength{\belowdisplayshortskip}{0pt}
	  \setlength{\intextsep}{1pt plus 1pt minus 1pt}
	  \setlength{\textfloatsep}{1pt plus 1pt minus 1pt}
	  \setlength{\skip\footins}{4pt plus 1pt minus 1pt}}
	\makeatother

\newcommand{\head}[1]{{\noindent\textbf{#1.}\xspace}} 

\newcolumntype{?}{!{\vrule width 1pt}} 
\newcolumntype{;}{!{\vrule width 0.5pt}}
\newcolumntype{P}[1]{>{\centering\arraybackslash}p{#1}}

\newcommand\ltitle{RUBICON: A Framework for Designing Efficient \\Deep Learning-Based  Genomic Basecallers}
\usepackage{fancyhdr}
\usepackage{datetime}
\author{Gagandeep Singh$^{a,c}$ \hspace{0.5cm}  Mohammed Alser$^{a}$ \hspace{0.5cm}    Kristof Denolf$^c$ \\  Can Firtina$^{a,*}$   \hspace{0.5cm} Alireza Khodamoradi$^{c}$ \hspace{0.5cm}  Meryem Banu Cavlak$^a$ \\ Henk Corporaal$^b$ \hspace{0.5cm} Onur Mutlu$^{a,*}$
\\ \normalsize$^a$Department of Information Technology and Electrical Engineering, ETH Z{\"u}rich, Switzerland  \\\normalsize $^b$Department of Electrical Engineering, Eindhoven University of Technology, The Netherlands \\\normalsize $^c$Research and Advanced Development, AMD, USA}

\newcommand{\hgb}[1]{{\color{black}{#1}}}
\newcommand{\hy}[1]{{\color{black}{#1}}}
\newcommand{\yboxbegin} {
	\begin{tcolorbox}[enhanced, frame hidden, colback=yellow!50, breakable]
}

\newcommand{\yboxend} {
	\end{tcolorbox}
}

\newcommand{\gbwriting}[1]{{\color{black}{#1}}}
\newcommand{\rmpSpeedupCC}{{364.89}$\times$\xspace}
\newcommand{\rmpSpeedupCRF}{{14.25}$\times$\xspace}
\newcommand{\rmpSpeedupBonCTC}{{128.13}$\times$\xspace}
\newcommand{\rmpSpeedupSAC}{{81.58}$\times$\xspace}
\newcommand{\rmpSpeedupDor}{{3.77}$\times$\xspace}
\newcommand{\rmpSpeedupFP}{{50.15}$\times$\xspace}
\newcommand{\rmpSpeedupSUP}{{185.54}$\times$\xspace}

\newcommand{\rfpSpeedupCC}{{7.28}$\times$\xspace}
\newcommand{\rfpSpeedupBonCTC}{{2.56}$\times$\xspace}
\newcommand{\rfpSpeedupSAC}{{1.63}$\times$\xspace}

\newcommand{\rmpAccCC}{{5.23}\%\xspace}
\newcommand{\rmpAccDor}{{2.85}\%\xspace}
\newcommand{\rmpAccCRF}{{2.89}\%\xspace}
\newcommand{\rmpAccBonCTC}{{0.06}\%\xspace}
\newcommand{\rmpAccSAC}{{1.97}\%\xspace}

\newcommand{\rmpAccSUP}{{2.47}\%\xspace}

\newcolumntype{y}{>{\columncolor{yellow}}r}

\newcommand{\boxbegin} {
	\begin{tcolorbox}[enhanced, frame hidden, colback=gray!50, breakable]
}
\newcommand{\boxend} {
	\end{tcolorbox}
}

\newcommand{\ogb}[1]{{\color{black}{#1}}}

\newcommand{\fgb}[1]{{\color{black}{#1}}}

\begin{document}
\sloppy
\title{\ltitle}

\maketitle
\def\thefootnote{*}\footnotetext{\fgb{Correspondence: firtinac@ethz.ch; omutlu@ethz.ch}}\def\thefootnote{\arabic{footnote}}
\section*{Abstract}
\begin{abstract}

\gbwriting{
Nanopore sequencing generates noisy electrical signals that need to be converted into a standard string of DNA nucleotide bases using a computational step called basecalling.
The performance of basecalling has critical implications for all later steps in genome analysis. 
Therefore, there is a need to reduce the computation and memory cost of basecalling while maintaining accuracy.
We present \framework, a framework to develop efficient hardware-optimized basecallers. 
We demonstrate the effectiveness of \framework by developing \mech, the first hardware-optimized mixed-precision basecaller that performs efficient basecalling, outperforming the state-of-the-art
basecallers. 
We believe \framework offers a promising
path to develop future hardware-optimized basecallers.}

\end{abstract}

\section*{Keywords}
genomics sequencing, basecalling, hardware acceleration, machine learning, deep neural network
\section{Background} 
\label{sec:introduction}
The rapid advancement of genomics and sequencing technologies continuously calls for the adjustment of existing algorithmic techniques or the development of entirely new computational methods across diverse biomedical domains~\cite{ginsburg2018precision, aryan2020moving, clark2019diagnosis, kingsmore2022genome,ginsburg2009genomic, bloom2021massively, quick2016realtime,yelagandula2021multiplexed, le2013selected, vladyslav2016whole, meyer2021critical, lapierre2020metalign, lapierre2019micop, meyer2022critical}. Modern sequencing machines~\cite{pollard_long_2018,senol_cali_nanopore_2019} are capable of sequencing complex genomic structures and variants with high accuracy and throughput using long-read sequencing technology~\cite{amarasinghe2020opportunities}. 
Oxford Nanopore Technologies (ONT) is the most widely used long-read sequencing technology~\cite{amarasinghe2020opportunities,logsdon2020long,wang2021nanopore, jain2018nanopore, gong2019ultralong,branton2008potential}. ONT devices generate long \sr{genomic} reads\sr{, each of which has} a length ranging from a few hundred to a million base pairs or nucleotides, i.e., A, C, G, and T in the DNA alphabet~\cite{van2018third,ardui2018single,jain_nanopore_2018,kchouk2017generations,weirather2017comprehensive}.

ONT devices sequence a genome by measuring changes to an electrical signal  as a single strand of DNA is passed through a nanoscale hole or \textit{nanopore}~\cite{jain2016oxford}. 
The generated \sr{noisy} electrical signal or \emph{squiggle} is decoded into a sequence of nucleotides using a computationally-expensive step\sr{, called} \emph{basecalling}~\cite{wick2019performance,pages2022comprehensive,wang2021nanopore,alser2022molecules, alser_technology_2021}. 
Basecallers need to address two key challenges to accurately basecall a raw sequencing input. First, provid\sr{ing} accurate predictions of each and every individual nucleotide, as the sensors measuring the changes in electrical current can only measure the effect of multiple neighboring nucleotides together~\cite{wick2019performance}. 
Second, tolerat\sr{ing} low signal-to-noise ratio (SNR) caused by thermal noise and the lack of statistically significant current signals triggered by DNA strand motions~\cite{pages2022comprehensive}.

Modern basecallers use deep learning-based models to significantly \sr{(by at least 10\%)} improve the accuracy \sr{of predicting a nucleotide base from the squiggle} compared to traditional non-deep learning-based basecallers~\cite{urnet_zhang2020nanopore,dias2019artificial, amarasinghe2020opportunities, firtina_apollo_2020, logsdon2020long, senol_cali_nanopore_2019, rang2018squiggle,alser2022molecules,mao2022genpip}. 
The success of deep learning in genome basecalling is attributed to the advances in its architecture to model and identify spatial features in raw input data to predict nucleotides.  
However, we observe the following six shortcomings with the current basecallers~\cite{urnet_zhang2020nanopore,catcaller_lv2020end,zeng2020causalcall,perevsini2021nanopore,lou2020helix,xu2021fast,konishi2021halcyon,huang2020sacall,neumann2022rodan}.
First, current state-of-the-art basecallers are slow and show poor performance on state-of-the-art CPU and GPU-based systems, bottlenecking the entire genomic analyses. 
\Copy{IR1.2/1}{\hgb{For example, state-of-the-art throughput optimized basecaller, \dor, takes $\sim$2.1 hours to basecall a 300 Gbps (Giga basepairs) human genome at 3$\times$ coverage on a server-grade GPU (NVIDIA A10G~\cite{a10} GPU with 24GiB DRAM and 16$\times$ CPU with 64 GiB DRAM)\mbox{\cite{guppydoradaBenchmark}}, while the subsequent step, i.e., read mapping, takes \ogb{only} a \ogb{small} fraction of basecalling time ($\sim$0.11 hours using minimap2\mbox{\cite{li_minimap_2016}}).}} 
\Copy{IR1.1/2}{\ogb{We observe that basecalling is the single longest stage in the genome sequencing pipeline, taking up to 43\% of execution time while the subsequent step of overlap finding, assembly, read mapping, and polishing take 18\%, 4\%, <1\%, and 35\% of execution time, respectively.}}

\Copy{IR1.1/1}{
\hgb{Second, for real-time sequencing, high basecalling throughput is a critical factor\mbox{\cite{quick2016realtime}}. In particular, scenarios such as \mbox{\emph{field sequencing}}\mbox{\cite{perevsini2021nanopore}} and \mbox{\emph{adaptive sampling}}\mbox{\cite{ulrich2022readbouncer}} necessitate rapid basecalling due to hardware limitations and the need for real-time decision-making. Field sequencing, often conducted in remote or resource-constrained environments, demands immediate basecalling to obtain actionable genomic information swiftly. Conventional high-compute infrastructure is often unavailable or impractical in these settings, underscoring the importance of an efficient basecalling process. Similarly, adaptive sampling protocols, aiming to optimize sequencing output based on real-time analysis of initial sequencing data, require a fast and accurate basecaller to make prompt decisions regarding read continuation or rejection. Also, enhancing the speed and efficiency of basecalling is critical for re-basecalling existing datasets using advanced, higher-accuracy models. By revisiting earlier data with improved basecalling algorithms, researchers can achieve a more precise representation of the genomic sequence. Current basecallers provide a tradeoff between speed and accuracy, often leading to sub-optimal performance in real-time sequencing scenarios.}}

Third, since basecalling shares similarities with automatic-speech recognition (ASR) task, many researchers have directly \sr{adapted} established ASR models, such as Quartznet~\cite{kriman2020quartznet}, Citrinet~\cite{majumdar2021citrinet}, and  Conformers~\cite{gulati2020conformer}, \sr{for basecalling} without customizing the neural network architecture specifically for the basecalling problem. 
Such an approach might lead to higher basecalling accuracy but at the cost of large and unoptimized neural network architecture. For example, \bon, 
an expert-designed convolutional neural network (CNN)-based version of  \texttt{Bonito} from ONT, has $\sim$10 million model parameters.    
We show in  Section~\ref{suppsubsubsec:overprovision_prune} that we can \sr{eliminate} up to 85\% of the model parameters to achieve a 6.67$\times$ reduction in model size without any loss in basecalling accuracy. Therefore, current basecalling models are costly to run, and the inference latency becomes a \sr{major} bottleneck.

Fourth, modern basecallers are typically composed of convolution layers with skip connections\footnote{A skip connection allows to skip some of the layers in the neural network and feeds the output of one layer as the input to the next layers.}~\cite{szegedy2017inception} (allow reusing 
of activations from previous layers)
that creates two major performance \sr{issues}: 
(a) skip connections increase the data lifetime: the layers whose activations are reused in future layers must either wait for this reuse to occur before accepting new input or store the activations for later use by utilizing more memory. Thus, leading to high resource and storage requirements; and (b) skip connections often need to perform additional computation to match the channel size at the input of the non-consecutive layer, which increases the number of model parameters; e.g., \bon requires $\sim$21.7\% additional model parameters due to the skip connections.
 
Fifth, current basecallers use floating-point precision (32 bits) to represent each neural network layer present in a basecaller. 
This leads to high bandwidth and processing demands~\cite{singh2019low,singh2020nero,singh2022designing}. Thus, current basecallers with floating-point arithmetic precision have inefficient hardware implementations. We observe in Section~\ref{suppsubsubsec:overprovision_quant} that the arithmetic precision requirements of current basecallers can be reduced $\sim$4$\times$ by adjusting the precision for each neural network layer based on the target hardware and desired accuracy. 

Sixth, basecallers that provide higher throughput have lower basecalling accuracy. For example, we show in Section~\ref{subsection:results_overall} and \fgb{Additional file 1:} Section S4  that \gpf provides up to 51.65$\times$ higher basecalling performance using  36.96$\times$ fewer model parameters at the expense of the   5.37\% lower basecalling accuracy compared to most accurate basecaller.

These six problems concurrently make basecalling slow, inefficient, and memory-hungry, bottlenecking all genomic analyses that depend on it. 
Therefore, there is a need to reduce the computation and memory cost of basecalling while maintaining their performance. However, developing a 
basecaller that can provide fast runtime performance with high accuracy requires a deep understanding of genome sequencing, machine learning, and hardware design. At present, computational biologists spend significant time and effort to design and implement new basecallers by an extensive trial-and-error process. 




\gsss{
\textbf{Our goal} is to overcome the above issues by developing a comprehensive framework for specializing and optimizing 
a deep learning-based basecaller that provides high efficiency and performance.}



\begin{figure*}[h]
  \centering
  \includegraphics[width=0.7\linewidth,bb=0 13 568 158]{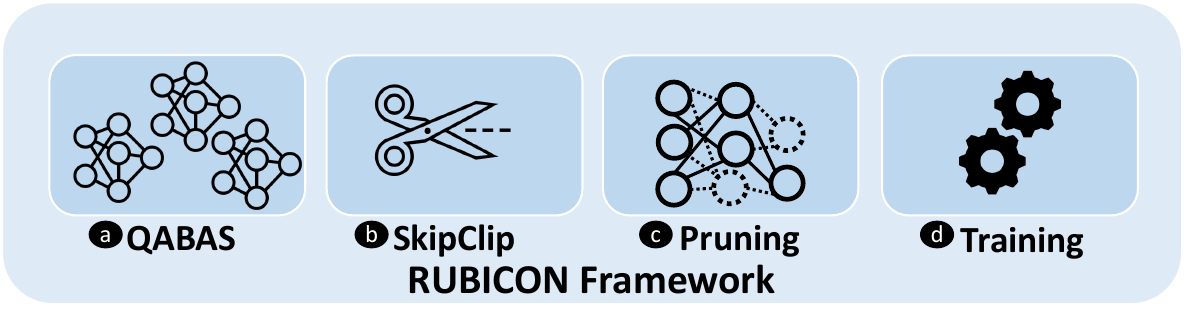}
  \caption{Overview of \framework framework. }
  \label{fig:framework}
\end{figure*}

To this end, we introduce \framework, the first framework for specializing and optimizing a machine learning-based basecaller. \framework uses  two machine learning techniques to develop hardware-optimized basecallers that are specifically designed for basecalling. First, we propose \nas, a quantization-aware basecalling architecture search framework to specialize basecaller architectures for hardware implementation while considering hardware performance metrics (e.g., latency, throughput, etc.). \nas uses neural architecture search (NAS)~\cite{zoph2016neural} to evaluate millions of different basecaller architectures. As discussed in \fgb{Additional file 1:} Section S1, during the basecaller neural architecture search, \nas quantizes the neural network model by exploring and finding the best bit-width precision for each neural network layer, which largely reduces the memory and computational complexity of a basecaller. 
 Adding quantization to the basecaller neural architecture search dramatically increases the model search space ($\sim$6.72$\times$10$^{20}$ more viable options in our search space). However, \sr{jointly} optimizing basecalling neural network architecture search and quantization allows us to develop accurate basecaller \sr{architectures} that are optimized for hardware acceleration. Second,
we develop \strim to remove all the skip connections \sr{present} in modern basecallers to reduce resource and storage requirements without any loss in basecalling
accuracy. \strim performs a skip removal process using knowledge distillation~\cite{bucilua2006model}, as shown in \fgb{Additional file 1:} Figure S2 in \fgb{Additional file 1:} Section S2, where  we train a smaller network (\textit{student}) without skip connections to mimic a pre-trained larger network (\textit{teacher}) with skip connections. \gsss{Figure~\ref{fig:framework} shows the key components of \framework. It consists
of four modules. \nas (\circled{a}) and \strim (\circled{b}) are two novel  techniques that are specifically designed for specializing and optimizing machine learning-based basecallers.  \framework provides support for \texttt{Pruning} (\circled{c}), which is a popular model compression technique where we discard network connections that are unimportant to neural network performance~\cite{lecun1989optimal,han2015deep,han2015learning,frankle2018lottery}. We integrate \texttt{Training} (\circled{d}) module from the official ONT basecalling pipeline~\cite{bonito}. For both the \texttt{Pruning} and \texttt{Training} modules, we provide the capability to use knowledge distillation~\cite{bucilua2006model,hinton2015distilling} for faster convergence and to increase the accuracy of the designed basecalling network.} 

\hfill \break
\head{Key results} 
We demonstrate the effectiveness of \framework by developing  \mech, the first hardware-optimized mixed-precision
basecaller that performs efficient basecalling, \hc{outperforming the state-of-the-art basecallers.} \fgb{Additional file 1:} Figure S5 in \fgb{Additional file 1:}  Section S2 shows the \mech architecture.  We compare \mech to five different basecallers. We demonstrate \gont{six} key results.  First,   \mech provides, on average, \rmpAccDor higher basecalling accuracy with \rmpSpeedupDor higher basecalling throughput compared to the fastest basecaller. Compared to 
an expert-designed basecaller \mech provides \rmpSpeedupBonCTC higher basecalling throughput without any loss in basecalling accuracy  by leveraging mixed precision computation when implemented on a cutting-edge spatial vector computing system, i.e., the AMD-Xilinx Versal AIE-ML~\cite{aiemlvc2802}. Second, we show that \nas-designed models are 5.74$\times$ smaller in size with 2.41$\times$ fewer neural network model parameters than an expert-designed basecaller. Third, by further using our \strim approach, \mech achieves a 6.88$\times$ and 2.94$\times$ reduction in neural network model size and the number of parameters, respectively. \Copy{IR1.5}{\gont{
Fourth, we show in \fgb{Additional file 1:} Section S4 that compared to the most accurate state-of-the-art basecaller \hgb{(i.e., \gp)}, \mech provides \rmpSpeedupSUP speedup using 19.22$\times$ lower parameters at the expense of, on average, \rmpAccSUP lower accuracy.}}  Fifth, assemblies constructed
using reads basecalled by \mech lead to higher quality, more contiguous, and more complete assemblies for all evaluated species than that provided by other basecallers. Sixth, \mech provides a 1.82\%-26.49\% lower number of base mismatches with the largest number of mapped bases and mapped reads compared to the baseline basecaller. Our experimental results on state-of-the-art computing systems show that \mech is a fast, memory-efficient, and hardware-friendly basecaller.  \framework can help researchers develop hardware-optimized basecallers that are superior to expert-designed models and can inspire independent future ideas.

\section{Results}
\label{sec:results}
\subsection{Analyzing the state-of-the-art basecaller} \label{suppsec:overprovision}

We observe established automatic-speech recognition (ASR) models being directly
applied to basecalling without optimizing it for basecalling. Such an approach leads to large and
unoptimized basecaller architectures. We evaluate the effect of using  two popular model compression techniques on the \bon basecaller: (1) Pruning, and (2) Quantization.

\subsubsection{Effect of pruning}
\label{suppsubsubsec:overprovision_prune}
\begin{figure}[h]
  \centering
  \includegraphics[width=1\linewidth,trim={0.2cm 0.85cm 0.4cm 0.3cm},clip]
  {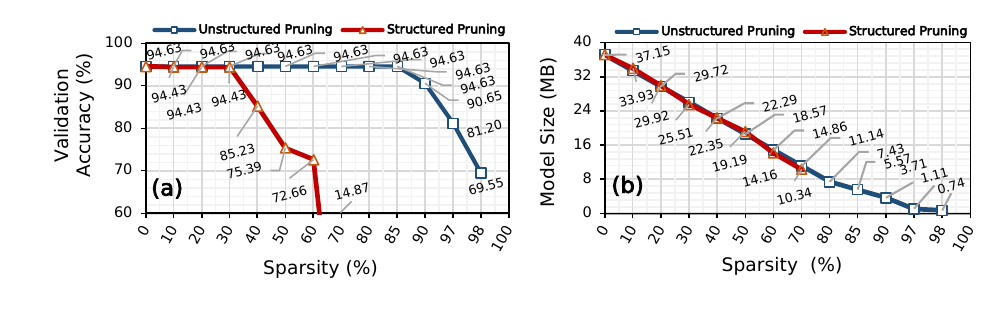}
  \caption{Effect of pruning the elements and channels of \bon using unstructured and structured pruning, respectively, on: (a) validation accuracy and (b) model size.}
  \label{supfig:over_prune}
\end{figure}

We show the effect of pruning \bon on the validation accuracy and model size in Figure~\ref{supfig:over_prune}(a) and Figure~\ref{supfig:over_prune}(b), respectively. Pruning is a model compression technique where we discard network connections that are unimportant to network performance without affecting the inference accuracy~\cite{lecun1989optimal,han2015deep,han2015learning,frankle2018lottery}.  We use unstructured element pruning and structured channel pruning with different degrees of sparsity. Unstructured or element pruning is a fine-grain way of pruning individual weights in a neural network without applying any pruning constraints. While in structured pruning, we remove a larger set of weights while maintaining a dense structure of the model~\cite{kruschke1991benefits, liu2018rethinking}.   

We make three major observations. First, \sr{pruning up to 85\% of the \bon model weights using unstructured pruning reduces the model size by 6.67$\times$ while maintaining the same accuracy as the baseline, unpruned \bon model.} Unstructured pruning leads to the highest model compression~\cite{gale2019state} at the cost of having sparse weights structure that is unsuitable for acceleration on any hardware platform.  While  pruning 30-40\% of the \bon model filters, using structured pruning reduces the model size by 1.46-1.66$\times$ while maintaining the same accuracy of the baseline, unpruned \bon model. Such \sr{a} high pruning ratio shows that most of the weights are redundant and \sr{do} not contribute to the actual accuracy. Second, after pruning 97\% (60\%) of the model weights, \bon provides 81.20\% (72.66\%) basecalling accuracy while using 33.33$\times$ (2.62$\times$) smaller model using unstructured pruning (structured pruning). Third, the \emph{knee point}\footnote{We define knee point as the point beyond which a basecaller is unable to basecall at an acceptable level of accuracy.} for unstructured pruning and structured pruning is at 98\% and 60\% where \bon provides 65.14\% and 72.66\% of basecalling accuracy, respectively. Beyond the knee-point, \bon losses its complete prediction power. We conclude that \bon is over-parameterized and \sr{contains} redundant logic and features.

\subsubsection{Effect of quantization}
\label{suppsubsubsec:overprovision_quant}
\begin{figure}[h]
  \centering
  \includegraphics[width=0.9\linewidth,trim={0.2cm 0.25cm 0.4cm 0.3cm},clip]{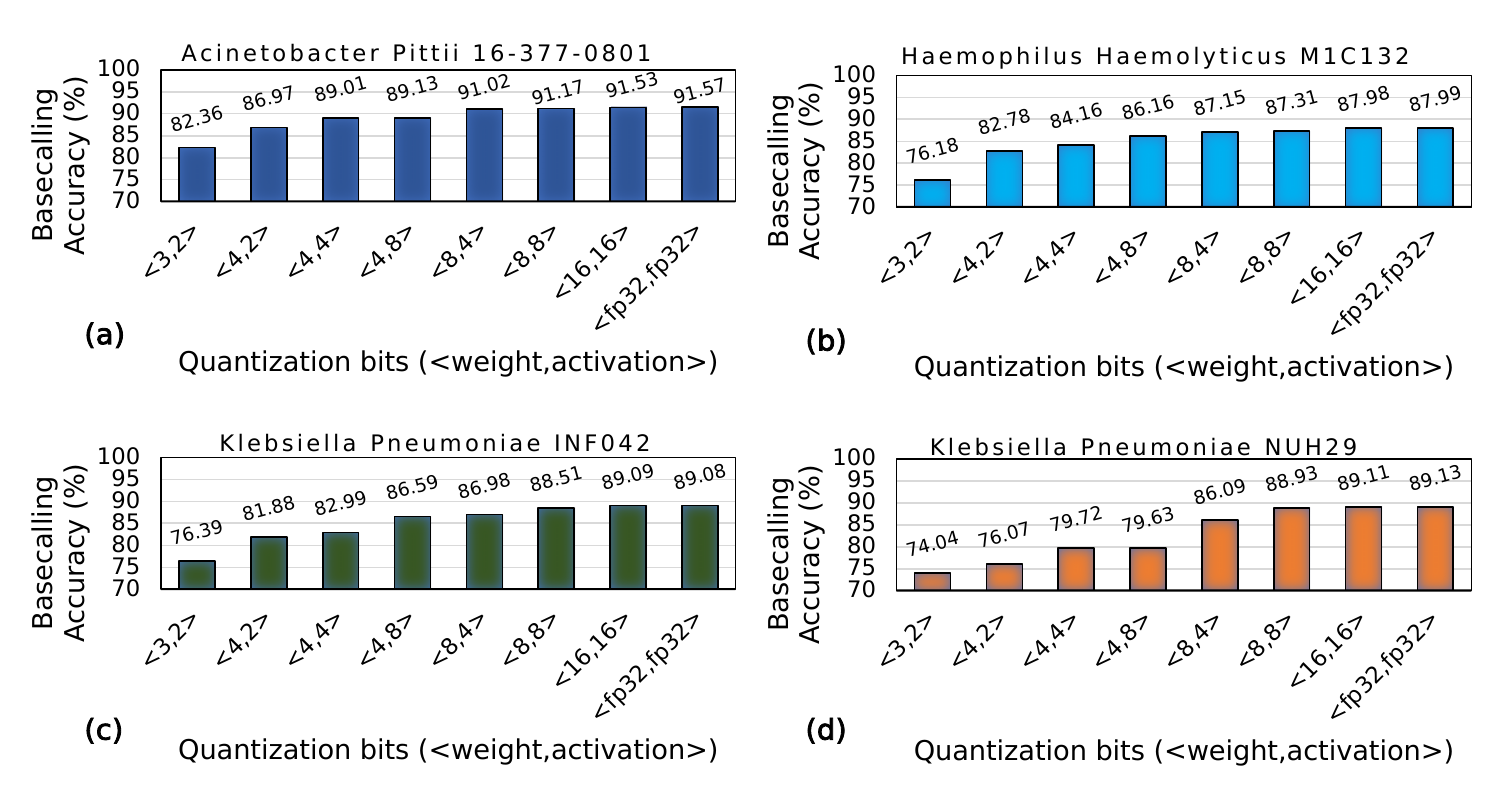}
  \vspace{-5pt}
  \caption{Basecalling using quantized models.}
  \label{supfig:over_quant}
\end{figure}

\begin{SCfigure}[][h]
  \centering
 \includegraphics[width=0.45\linewidth,trim={0cm 0cm 0.2cm 0.35cm},clip]{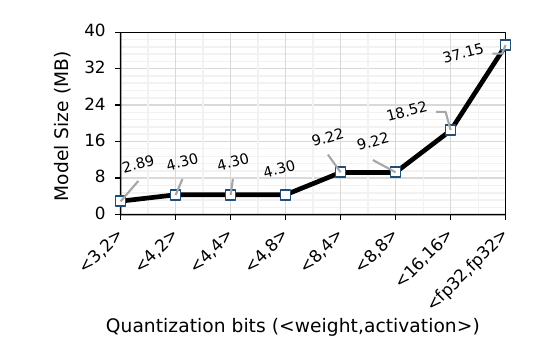}
 \vspace{-0.2cm}
 \hspace{-0.6cm} \caption{\protect\rule{0ex}{5ex}Effect of quantizing weight and activation of \bon on model size. We quantize both the weight and activation with static precision. Since weights are the trainable parameters in a neural network,  only weights contribute to the final model size.}
\label{supfig:quant_def_model_size}
\end{SCfigure}

Figure~\ref{supfig:over_quant} shows the effect \sr{of} using a quantized model to basecall \gsss{on the basecalling accuracy \hc{for four different species}.} 
In Figure~\ref{supfig:quant_def_model_size}, we show the effect of quantization on the model size. We quantize both the weight and activation using six different bit-width configurations ($<$\texttt{3,2}$>$, $<$\texttt{4,2}$>$, $<$\texttt{4,4}$>$, $<$\texttt{4,8}$>$, $<$\texttt{8,4}$>$, and $<$\texttt{16,16}$>$). 
We also show the results with the default floating-point precision ($<$\texttt{fp32,fp32}$>$). We use static quantization that uses the same precision for each neural network layer.

  \vspace{5pt}

We make four main observations. First, \sr{using a precision of $<$\texttt{8,8}$>$ for weight and activation for all the layers \sr{of} \bon causes a negligible accuracy} loss (0.18\%-0.67\%), while \sr{reducing the model size by} 4.03$\times$. Second, \bon is more sensitive to weight precision than activation precision.
\sr{For example, we} observe a loss of 1.82\%-9.48\% accuracy \sr{when using a precision of} $<$\texttt{4,8}$>$ instead of $<$\texttt{16,16}$>$ bits compared to an accuracy loss of only 0.51\%-3.02\% \sr{when using a precision of} $<$\texttt{8,4}$>$ instead of $<$\texttt{16,16}$>$ bits. 
Third, we observe a significant drop in accuracy (\sr{by} 9.17\%-15.07\%), when using \sr{less} than 4 bits for weights (e.g., using $<$\texttt{3,2}$>$ configuration). Fourth, \sr{using bit-width precision of $<$16,16$>$ bits provides $\sim$2$\times$ reductions in model size and without any \sr{accuracy} loss compared to using full precision ($<$\texttt{fp32,fp32}$>$) floating\sr{-point} implementation.} 
We conclude that the current state-of-the-art basecaller, \bon, can \sr{still} efficiently \sr{perform basecalling even when using} lower precision for both the weight and activation.

\subsection{\mech: Overall trend}
\label{subsection:results_overall}
\begin{figure}[H]  
  \centering
  \includegraphics[width=1\linewidth]{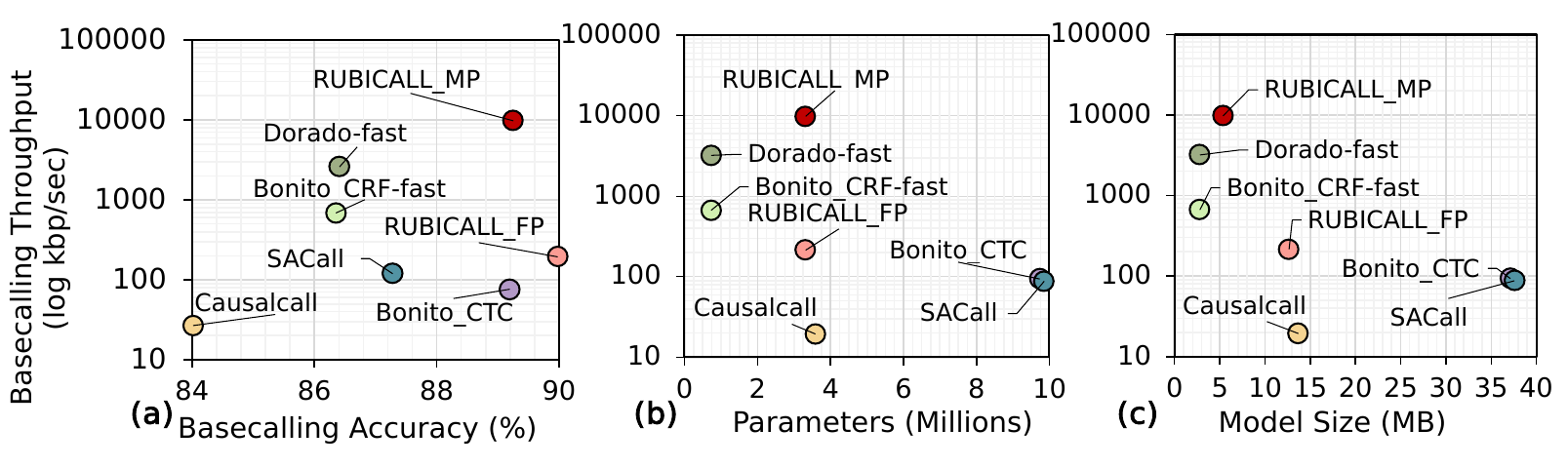}
  \caption{Comparison of \hc{average} basecalling  throughput for \mechmp with state-of-the-art basecallers in terms of:  (a) average  basecalling accuracy, (b) model parameters, and (c) model size. \mechmp provides higher compute performance with lower model size when compared to \mechfp because of the mixed-precision computation.}
  \label{fig:model_compare}
\end{figure}

\sr{We compare} the \sr{overall} basecalling \sr{throughput} \sr{of} \mech with \sr{that of the} baseline basecaller\sr{s} in terms of average basecalling accuracy, model parameters, and model size in Figure~\ref{fig:model_compare}(a), \ref{fig:model_compare}(b), and \ref{fig:model_compare}(c), respectively. We evaluate \mech using: (1) MI210 GPU~\cite{mi210} (\mech-FP) \gssss{using floating-point precision computation}, and (2) Versal ACAP VC2802~\cite{aiemlvc2802}, a  cutting-edge spatial vector computing system (\mech-MP) \gssss{using mixed-precision computation}. \gont{Section~\ref{sec:evaluation} provides details on our evaluation methodology.}

\sr{We make six key observations.}
First, \hy{compared to \dor, the fastest basecaller, \mbox{\mechmp} provides, on average, \rmpAccDor higher accuracy with \rmpSpeedupDor higher basecalling throughput. Therefore, \mbox{\mechmp} provides both accuracy and high basecalling throughput.
Second,  \mechmp provides \rmpSpeedupBonCTC higher basecalling throughput without any loss in accuracy compared to \bon, which is an expert-designed basecaller. Unlike \bon, This is because \mechmp has a mixed precision neural architecture that leads to high compute density. 
Third, by using mixed-precision quantization, \texttt{RUBICALL}-\texttt{MP} provides \rmpSpeedupFP higher performance  when compared to its floating-point implementation (\mechfp).
Fourth, \mbox{\sac} has the highest number of neural network model parameters, which are 2.74$\times$, 13.49$\times$, 1.01$\times$, 13.49$\times$, and 2.97$\times$ more than \cc, \gpf, \bon, \dor, and \mechmp, respectively. \sac uses a large transformer model with an attention mechanism that leads to an over-parameterized model.} 
Fifth, \dor has 4.92$\times$, 13.33$\times$, 13.49$\times$,  and 4.54$\times$ lower number of trainable \sr{model} parameters than \cc, \bon, \sac, and \mechmp. As discussed earlier, \dor provides \rmpAccDor lower accuracy with \rmpSpeedupDor lower basecalling throughput. While \dor has a 4.54$\times$ lower number of trainable \sr{model} parameters, the difference in model size is only 1.92$\times$ because \mechmp has each layer quantized to a different precision.
Sixth, compared to basecallers with skip connections, \mechmp provides 2.55$\times$ and \sr{6.93$\times$} smaller \sr{model} size compared to \cc and \bon, respectively. 
The decrease in model size is due to: (1) \sr{a} lower number of neural network layers; and (2) optimum bit-width precision for each neural network layer. Sixth, all the baseline basecallers use floating-point arithmetic precision for all neural network layers. This leads to very high memory bandwidth and processing demands. We conclude that \mechmp provides the ability to basecall quickly, and efficiently scale basecalling by providing reductions in both model size and  neural network model parameters.

\subsection{Performance comparison}
\label{subsection:results_perf_compare}
\begin{figure}[H]
  \centering
  \includegraphics[width=0.8\linewidth]{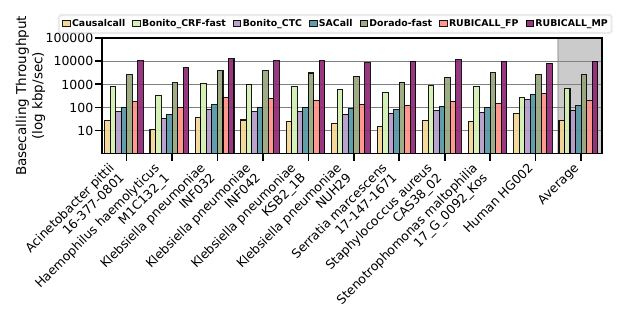}
  \vspace{-13pt}
  \caption{Performance comparison of \mech (using floating-point precision (\mechfp) and mixed-precision (\mechmp))~\sr{and five state-of-the-art basecallers on AMD MI210. The y-axis is on a logarithmic scale.}}
  \label{fig:perf}
  \vspace{12pt}
\end{figure}

\sr{We compare} the speed of \mechmp against baseline basecallers in Figure~\ref{fig:perf}.
We make three major observations. First, \mechmp consistently outperforms all the other basecallers for all the evaluated species. 
\mechmp improves average performance by \rmpSpeedupCC, \rmpSpeedupCRF, \rmpSpeedupBonCTC, \rmpSpeedupSAC, and \rmpSpeedupDor over \cc,  \gpf, \bon, \sac, and \mbox{\dor}, respectively. Second, as \texttt{RUBICALL}-\texttt{MP} each layer is quantized to a different precision, it provides \rmpSpeedupFP higher performance when compared to its floating-point only implementation (\mechfp).
Third, \mechfp, by using floating-point precision, provides \rfpSpeedupCC, \rfpSpeedupBonCTC, and \rfpSpeedupSAC higher performance compared to \cc, \bon, and \sac, respectively.  \Copy{CC1.2/2}{\hgb{\fgb{Additional file 1:} Figure S7 in \fgb{Additional file 1:}  Section S5 demonstrates the performance of all the evaluated basecallers on NVIDIA A40~\cite{a40} GPU.}}   We conclude that using mixed-precision computation, \mechmp consistently performs better than the baseline basecallers.

\subsection{Basecalling accuracy}
\label{subsection:results_base_acc}
We compare the basecalling accuracy of \mech against baseline basecallers in Figure \ref{fig:identity}.  \mechmp and \mechfp use the same model architecture and produce the same basecalled reads, so we report results as \mech.
We make three major observations. 
First, compared to \dor and \gpf, we observe \mbox{\mech} achieves \rmpAccDor and \rmpAccCRF higher accuracy over these RNN-based basecallers, respectively. \mech provides \rmpAccCC and \rmpAccBonCTC higher accuracy than CNN-based basecaller \cc and \bon, respectively. Compared to a state-of-the-art transformer-based basecaller, \sac, \mbox{\mech} achieves \rmpAccSAC higher basecalling accuracy.  Second, \bon has 2.93$\times$ higher parameters (Figure~\ref{fig:model_compare}(a)) while having similar accuracy as \mech. 
Third, \cc and \sac are unable to align half of \emph{Haemophilus haemolyticus M1C132\_1} reads to its reference. Therefore, it is deemed unaligned and cannot be used to determine its read accuracy. 
We conclude that \gs{\mech provides the highest accuracy compared to other basecallers.}

\begin{figure}[H]
  \centering
  \includegraphics[width=0.8\linewidth]{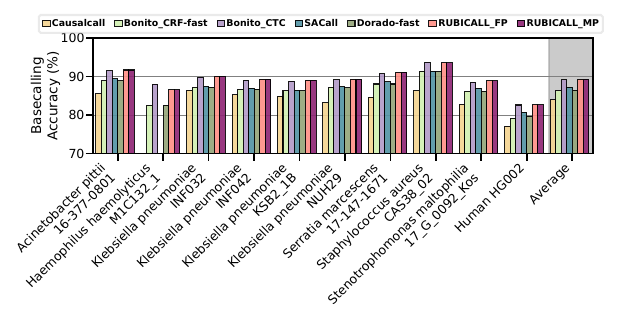} 
  \vspace{-15pt}
  \caption{\gs{Basecalling accuracy} comparison of \mech (using floating-point precision (\mechfp) and mixed-precision (\mechmp)).}
  \label{fig:identity}
\end{figure}


\subsection{Downstream analysis}
\subsubsection{De novo assembly}
\label{subsubsection:down_denovo}
\hy{We provide the statistics related to the accuracy, completeness, and contiguity of assemblies we generate using the basecalled reads from \cc, \gpf, \bon, \sac, \dor, and \mbox{\mech} in \mbox{Table~\ref{tab:downstream}}.} \Copy{IR1.14}{\hgb{For \mbox{\texttt{Genome Fraction (\%)}}, \mbox{\texttt{Average Identity (\%)}}, and Quality Value (QV), we highlight the highest achieved value. While for \mbox{\texttt{Assembly Length}}, \mbox{\texttt{Average GC (\%)}}, and \texttt{NG50}, we highlight the value closest to the real assembly length.} \Copy{IR3.10/2}{\hgb{For \mbox{\texttt{Total Indels}} and \mbox{\texttt{Indel Ratio (\%)}}, the best-performing basecaller has the lowest value.}}}
\Copy{IR3.6/2}{\ogb{We also collect the number of unique k-mers and the frequency of each unique k-mer in a given sequence to perform a comparison of under and over-represented k-mers in \fgb{Additional file 1:}  Section S7.}}
\begin{table*}[htbp]
 \caption{Assembly quality comparison of the evaluated basecallers for different species. We measure assembly accuracy in terms of genome fraction (Genome Fraction (\%)) and average identity (Average Identity (\%)). Genome fraction is the portion of the \texttt{\texttt{Reference}} genome
that can align to a given assembly, while average identity is the average of the identity of assemblies when compared to their respective \texttt{\texttt{Reference}} genomes. We measure statistics related to the contiguity
and completeness of the assemblies in terms of the overall assembly length (Assembly Length), Average GC content (Average GC (\%)) (i.e., the ratio of G and C bases in an assembly),  NG50 statistics (NG50) (i.e., shortest contig at the half of the overall \texttt{\texttt{Reference}} genome length), \hgb{total number of indels in all aligned bases in the assembly (Total Indels), the ratio of indels to assembly length (Indel Ratio (\%)), and the reliability of basepairs using the quality value (Quality Value)}. NA indicates that the generated assemblies were unalignable to the reference genome.}
    \label{tab:downstream}
\centering
 \setstretch{0.8}
\renewcommand{\arraystretch}{1}
  \resizebox{0.85\linewidth}{!}{%
\begin{tabular}{llrrrrrrrr}
\hline
\textbf{Dataset}   & \textbf{Basecaller}   & \multicolumn{1}{l}{\textbf{\begin{tabular}[c]{@{}l@{}}Genome \\ Fraction (\%)\end{tabular}}} & \multicolumn{1}{l}{\textbf{\begin{tabular}[c]{@{}l@{}}Average \\ Identity (\%)\end{tabular}}} & \multicolumn{1}{l}{\textbf{\begin{tabular}[c]{@{}l@{}}Assembly \\ Length\end{tabular}}} & \multicolumn{1}{l}{\textbf{\begin{tabular}[c]{@{}l@{}}Average \\ GC (\%)\end{tabular}}} & \multicolumn{1}{l}{\textbf{NG50}} & \multicolumn{1}{l}{\textbf{\begin{tabular}[c]{@{}l@{}}Total \\ Indels\end{tabular}}} & \multicolumn{1}{l}{\textbf{\begin{tabular}[c]{@{}l@{}}Indel \\ Ratio (\%)\end{tabular}}} & \multicolumn{1}{l}{\textbf{\begin{tabular}[c]{@{}l@{}}Quality\\ Value (QV)\end{tabular}}} \\
\hline
Acinetobacter      & \cc            & 92.45                                                                                        & 86.18                                                                                         & \textbf{3,826,077}                                                                      & 42.23                                                                                   & \textbf{3,826,077}                & 270,228                                                                              & 7.06      &11.99                                                                               \\
pittii 16-377-0801 & \gpf & 96.64                                                                                        & 89.29                                                                                         & 3,628,317                                                                               & \textbf{38.82}                                                                        & 3,628,317                         & 242,373                                                                              & 6.68              &12.03                                                                         \\
& \bon          & \textbf{96.87}                                                                                        & 91.44                                                                                         & 3,676,821                                                                               & 38.9                                                                                    & 3,676,821                         & 210,496                                                                              & 5.72                  &12.45                                                                   \\
& \sac                & 96.68                                                                                        & 89.42                                                                                         & 3,699,232                                                                               & 38.7                                                                                    & 3,699,232                         & 247,997                                                                              & 6.7      &12.1                                                                                \\
& \dor                & 96.37                                                                                        & 88.72                                                                                         & 3,839,847                                                                               & 39.09                                                                                   & 3,839,847                         & 245,016                                                                              & 6.38            &12.03                                                                         \\
& \mech              & \textbf{96.87}                                                                               & \textbf{91.51}                                                                                & 3,694,086                                                                               & \textbf{38.82}                                                                          & 3,694,086                         & \textbf{208,748}                                                                     & \textbf{5.65}            &\textbf{15.42}                                                                \\
& \texttt{Reference}             & 100                                                                                          & 100                                                                                           & 3,814,719                                                                               & 38.78                                                                                   & 3,814,719                         & 0                                                                                    & 0      & -                                                                                  \\\hline
Haemophilus        & \cc            & 0.00                                                                                         & 0.00                                                                                          & 0                                                                                       & 0                                                                                       & 0                                 & 0                                                                                    & 0                                       &NA                                                 \\
haemolyticus       & \gpf & 88.76                                                                                        & \textbf{91.51}                                                                                & \textbf{2,046,024}                                                                               & 37.98                                                                                   & \textbf{2,046,024}                         & 128,481                                                                              & 6.28                                                         &12.25                            \\
M1C132\_1          & \bon          & \textbf{96.87 }                                                                                       & 90.70                                                                                         & 1,957,480                                                                               & 38.87                                                                                   & {1,957,480}                & \textbf{118,253}                                                                     & \textbf{6.04}            &15.34                                                                \\
& \sac                & 90.11                                                                                        & 88.45                                                                                         & 2,032,994                                                                               & \textbf{38.22}                                                                          & 1,880,730                         & 134,702                                                                              & 6.63                                        &13.15                                             \\
& \dor                & 89.42                                                                                        & 88.97                                                                                         & 2,110,860                                                                               & 39.49                                                                                   & 2,110,860                         & 129,503                                                                              & 6.14                                           &12.38                                          \\
& \mech              & \textbf{96.87}                                                                               & 90.54                                                                                         & {1,966,781}                                                                      & 38.92                                                                                   & 1,966,781                         & 119,777                                                                              & 6.09   &\textbf{15.37}                                                                                  \\
& \texttt{Reference}             & 100                                                                                          & 100                                                                                           & 2,042,591                                                                               & 38.46                                                                                   & 2,042,591                         & 0                                                                                    & 0       & -                                                                                 \\\hline
Klebsiella         & \cc            & 92.45                                                                                        & 87.35                                                                                         & \textbf{4,959,127}                                                                      & 56.9                                                                                    & 4,959,127                         & 353,550                                                                              & 7.13                       &10.54                                                              \\
pneumoniae         & \gpf & 92.69                                                                                        & 87.53                                                                                         & 4,761,297                                                                               & \textbf{57.19}                                                                          & 4,761,297                         & 347,299                                                                              & 7.29                         &10.56                                                            \\
INF032             & \bon          & 94.50                                                                                        & 90.20                                                                                         & 4,897,352                                                                               & 56.65                                                                                   & 4,897,352                         & 317,428                                                                              & 6.48        &11.26                                                                             \\
& \sac                & 93.97                                                                                        & 88.08                                                                                         & 4,874,880                                                                               & 56.87                                                                                   & 4,874,880                         & 379,028                                                                              & 7.78    &10.8                                                                                 \\
& \dor                & 93.00                                                                                        & 87.69                                                                                         & 5,063,562                                                                               & 56.8                                                                                    & \textbf{5,063,562}                & 348,572                                                                              & 6.88     &10.64                                                                                \\
& \mech              & \textbf{94.51}                                                                               & \textbf{90.30}                                                                                & 4,924,240                                                                               & 56.85                                                                                   & 4,924,240                         & \textbf{314,651}                                                                     & \textbf{6.39}                             & \textbf{11.27}                                               \\
& \texttt{Reference}             & 100                                                                                          & 100                                                                                           & 5,111,537                                                                               & 57.63                                                                                   & 5,111,537                         & 0                                                                                    & 0                                 & -                                                       \\\hline
Klebsiella         & \cc            & 91.44                                                                                        & 87.36                                                                                         & \textbf{5,288,166}                                                                      & \textbf{56.94}                                                                          & \textbf{5,288,166}                & 374,162                                                                              & 7.08   &10.84                                                                                  \\
pneumoniae         & \gpf & 92.08                                                                                        & 88.49                                                                                         & 5,052,889                                                                               & 56.8                                                                                    & 5,052,889                         & 357,354                                                                              & 7.07    & 10.93                                                                                 \\
INF042             & \bon          & 93.12                                                                                        & 90.49                                                                                         & 5,111,083                                                                               & 56.61                                                                                   & 5,111,083                         & 317,075                                                                              & 6.2       & 11.40                                                                               \\
& \sac                & 92.93                                                                                        & 88.60                                                                                         & 5,149,039                                                                               & 56.72                                                                                   & 5,149,039                         & 369,388                                                                              & 7.17         & 11.08                                                                            \\
& \dor                & 90.21                                                                                        & 88.20                                                                                         & 5,737,059                                                                               & 56.44                                                                                   & 5,401,717                         & 342,141                                                                              & 5.96    & 10.98                                                                                 \\
& \mech              & \textbf{93.12}                                                                               & \textbf{90.60}                                                                                & 5,146,050                                                                               & 56.72                                                                                   & 5,146,050                         & \textbf{312,448}                                                                     & \textbf{6.07} & \textbf{11.42}                                                                          \\
& \texttt{Reference}             & 100                                                                                          & 100                                                                                           & 5,337,491                                                                               & 57.41                                                                                   & 5,337,491                         & 0                                                                                    & 0     & -                                                                                   \\\hline
Klebsiella         & \cc            & 91.58                                                                                        & 86.97                                                                                         & \textbf{5,175,311}                                                                      & \textbf{57.09}                                                                          & 5,175,311                         & 363,807                                                                              & 7.03      & 10.88                                                                               \\
pneumoniae         & \gpf & 90.24                                                                                        & 88.00                                                                                         & 4,932,626                                                                               & 56.71                                                                                   & 4,932,626                         & 357,769                                                                              & 7.25                                  &10.86                                                   \\
KSB2\_1B           & \bon          & 93.07                                                                                        & \textbf{90.11}                                                                                & 5,003,377                                                                               & 56.69                                                                                   & 5,003,377                         & \textbf{320,519}                                                                     & \textbf{6.41}  &\textbf{11.41}                                                                          \\
& \sac                & \textbf{93.58}                                                                               & 88.19                                                                                         & 5,034,408                                                                               & 56.79                                                                                   & 5,034,408                         & 372,380                                                                              & 7.4                                                             & 11.16                         \\
& \dor                & 90.28                                                                                        & 87.67                                                                                         & 5,442,186                                                                               & 56.72                                                                                   & \textbf{5,261,731}                & 349,387                                                                              & 6.42      &11.03                                                                               \\
& \mech              & 93.07                                                                                        & 89.89                                                                                         & 5,023,639                                                                               & 56.75                                                                                   & 4,932,626                         & 357,769                                                                              & 7.12                                        &11.25                                             \\

& \texttt{Reference}             & 100                                                                                          & 100                                                                                           & 5,228,889                                                                               & 57.59                                                                                   & 5,228,889                         & 0                                                                                    & 0                                                                  & -                      \\\hline
Klebsiella         & \cc            & 89.08                                                                                        & 86.01                                                                                         & \textbf{5,158,874}                                                                      & 56.78                                                                                   & \textbf{5,158,874}                & 389,676                                                                              & 7.55      &11.75                                                                               \\
pneumoniae         & \gpf & 92.17                                                                                        & 89.34                                                                                         & 4,942,833                                                                               & 57.01                                                                                   & 4,942,833                         & 355,690                                                                              & 7.2                                                                        &11.47              \\
NUH29              & \bon          & \textbf{94.36}                                                                               & 90.26                                                                                         & 4,918,147                                                                               & 57.04                                                                                   & 4,918,147                         & 324,406                                                                              & 6.6                                              &\textbf{11.92}                                        \\
& \sac                & 93.66                                                                                        & 88.58                                                                                         & 4,978,307                                                                               & 57.06                                                                                   & 4,978,307                         & 360,950                                                                              & 7.25     &11.56                                                                                \\
& \dor                & 92.27                                                                                        & 88.12                                                                                         & 5,195,594                                                                               & 57.01                                                                                   & 5,195,594                         & 355,728                                                                              & 6.85       &11.56                                                                              \\
& \mech              & \textbf{94.36}                                                                               & \textbf{90.43}                                                                                & 4,940,813                                                                               & \textbf{57.18}                                                                          & 4,940,813                         & \textbf{316,019}                                                                     & \textbf{6.4}    &11.83                                                                         \\
& \texttt{Reference}             & 100                                                                                          & 100                                                                                           & 5,134,281                                                                               & 57.61                                                                                   & 5,134,281                         & 0                                                                                    & 0                                                       & -                                 \\\hline
Serratia           & \cc            & 89.91                                                                                        & 86.23                                                                                         & \textbf{5,532,953}                                                                      & 57.86                                                                                   & \textbf{5,422,052}                & 401,545                                                                              & 7.26     &\textbf{13.39}                                                                                \\
marcescens         & \gpf & 96.06                                                                                        & 89.56                                                                                         & 5,479,812                                                                               & \textbf{58.85}                                                                          & 5,282,474                         & 345,351                                                                              & 6.3                                     &12.66                                                 \\
17-147-1671        & \bon          & \textbf{96.76}                                                                               & 91.38                                                                                         & 5,534,329                                                                               & 58.41                                                                                   & 5,316,651                         & 298,982                                                                              & 5.4         &13                                                                             \\
& \sac                & 94.29                                                                                        & 89.36                                                                                         & 5,366,913                                                                               & 58.57                                                                                   & 5,366,913                         & 358,954                                                                              & 6.69     &12.27                                                                                \\
& \dor                & 96.51                                                                                        & 88.87                                                                                         & 5,758,989                                                                               & 58.29                                                                                   & 5,282,474                         & 348,968                                                                              & 6.06    &12.5                                                                                 \\
& \mech              & \textbf{96.76}                                                                               & \textbf{91.59}                                                                                & 5,597,251                                                                               & 58.52                                                                                   & 5,346,640                         & \textbf{294,643}                                                                     & \textbf{5.26} & 13.01                                                                           \\
& \texttt{Reference}             & 100                                                                                          & 100                                                                                           & 5,517,578                                                                               & 59.13                                                                                   & 5,517,578                         & 0                                                                                    & 0                                                         & -                               \\\hline
Staphylococcus     & \cc            & 94.35                                                                                        & 87.29                                                                                         & 2,849,123                                                                               & 36.59                                                                                   & 2,810,038                         & 191,730                                                                              & 6.73        & 10.8                                                                             \\
aureus             & \gpf & 96.27                                                                                        & 91.49                                                                                         & 2,790,895                                                                               & 33.05                                                                                   & 2,752,169                         & 149,623                                                                              & 5.36  & 11.59                                                                                   \\
CAS38\_02          & \bon          & \textbf{97.03}                                                                                       & \textbf{93.57}                                                                                & 2,858,986                                                                               & \textbf{32.86}                                                                          & 2,819,356                         & \textbf{123,542}                                                                     & \textbf{4.32}  & \textbf{12.82}                                                                          \\
& \sac                & 95.66                                                                                        & 91.25                                                                                         & 2,837,503                                                                               & 32.91                                                                                   & 2,798,079                         & 165,200                                                                              & 5.82  & 11.57                                                                                   \\
& \dor                & 96.70                                                                                        & 91.16                                                                                         & 2,927,882                                                                               & 33.52                                                                                   & 2,752,169                         & 152,216                                                                              & 5.2    & 11.64                                                                                  \\
& \mech              & \textbf{97.03}                                                                               & 93.36                                                                                         & \textbf{2,860,885}                                                                      & 33.24                                                                                   & \textbf{2,821,276}                & 124,795                                                                              & 4.36              & 12.59                                                                       \\
& \texttt{Reference}             & 100                                                                                          & 100                                                                                           & 2,902,076                                                                               & 32.82                                                                                   & 2,902,076                         & 0                                                                                    & 0                           & -                                                             \\\hline
Stenotrophomonas   & \cc            & 94.85                                                                                        & 85.73                                                                                         & \textbf{4,823,177}                                                                      & 63.66                                                                                   & \textbf{4,823,177}                & 366,228                                                                              & 7.59    & 11.01                                                                                 \\
maltophilia        & \gpf & 94.60                                                                                        & 89.74                                                                                         & 4,596,898                                                                               & \textbf{65.5}                                                                           & 4,596,898                         & 337,040                                                                              & 7.33                                                                    &11.10                 \\
17\_G\_0092\_Kos   & \bon          & 95.42                                                                                        & 90.14                                                                                         & 4,664,226                                                                               & 64.82                                                                                   & 4,664,226                         & 298,711                                                                              & 6.4        & 11.51                                                                              \\
& \sac                & 95.28                                                                                        & 88.50                                                                                         & 4,672,540                                                                               & 64.98                                                                                   & 4,672,540                         & 339,853                                                                              & 7.27   & 11.11                                                                                  \\
& \dor                & 92.99                                                                                        & 87.70                                                                                         & 4,854,007                                                                               & 63.99                                                                                   & 4,854,007                         & 337,105                                                                              & 6.94     &11.01                                                                                \\
& \mech              & \textbf{95.46}                                                                               & \textbf{90.49}                                                                                & 4,693,744                                                                               & 65.03                                                                                   & 4,693,744                         & \textbf{289,073}                                                                     & \textbf{6.16}  & \textbf{11.63}                                                                          \\
& \texttt{Reference}             & 100                                                                                          & 100                                                                                           & 4,802,733                                                                               & 66.28                                                                                   & 4,802,733                         & 0                                                                                    & 0                                                      & -                                  \\\hline

Human              & \cc            & NA                                                                       & NA                                                                                            & 130,962                                                                                 & 42.95                                                                                   & 13,522                            & NA                                                               & NA     & NA                                                              \\

HG002              & \gpf & 0.002                                                                                        & 92.36                                                                                         & 119,570,537                                                                             & 40.34                                                                                   & 368,848                           & 2,860                                                                                & \textbf{0}      & \textbf{18.87}                                                                         \\
 
& \bon          & \textbf{0.430}                                                                               & \textbf{95.06}                                                                                & 134,732,516                                                                             & \textbf{40.86}                                                                          & 371,590                           & 384,243                                                                              & 0.29        & 18.58                                                                             \\

& \sac                & NA                                                                       & NA                                                                                            & 63,025,520                                                                              & 39.87                                                                                   & 320,873                           & NA                                                               & NA     & NA                                                              \\

& \dor                & 0.001                                                                                        & 93.15                                                                                         & 121,146,376                                                                             & 39.8                                                                                    & 361,677                           & \textbf{926}                                                                         & \textbf{0}   & 17.46                                                                            \\

& \mech              & 0.125                                                                                        & 94.50                                                                                         & \textbf{140,928,248}                                                                    & 40.99                                                                                   & \textbf{393,950}                  & 100,256                                                                              & 0.1      & 17.81                                                                                \\

& \texttt{Reference}             & 100                                                                                          & 100                                                                                           & 2,947,743,500                                                                           & 40.79                                                                                   & 2,947,743,500                     & 0                                                                                    & 0                                                         & -                               \\\hline
                              
\end{tabular}
}
\end{table*}

We make six key observations. First, assemblies constructed \sr{using reads basecalled} by \mech provide the best reference genome coverage for \emph{all} datasets (``Genome Fraction'' in Table~\ref{tab:downstream}). 
This means that \sr{assemblies built using \mech-basecalled reads} \sr{are more complete than assemblies built using reads from other basecallers} \sr{since a larger portion of the corresponding reference genomes align to their assemblies using \mech-basecalled reads compared to that of using reads from other basecallers}. 
\hy{Second, assemblies constructed using the \mbox{\mech} reads  usually have a higher average identity than that of \cc, \gpf, \bon, \sac, and \dor.} 
These average identity results are tightly in line with the basecalling accuracy results we show in Figure~\ref{fig:identity}. 
\sr{Although} \gpf provides \sr{a} higher average identity for the Haemophilus haemolyticus M1C132\_1 dataset (i.e., 91.51\%), \sr{the} genome coverage \sr{provided by both} \gpf and \dor is 2.2\% lower than that \sr{provided by} \mech for the same dataset. \sr{This means a large} portion of the assembly \sr{provided by} \gpf ~\sr{has low-quality} regions as 
the reference genome cannot align to these regions due to high dissimilarity.
Third, \sr{assemblies constructed using the \mech reads} provide better completeness and contiguity as they have 1)~assembly lengths closer to their corresponding reference genomes and 2)~higher NG50 results in most cases than \sr{those constructed using the \gpf and \bon reads}. 
Fourth, although \cc usually provides the best results in terms of the assembly lengths and NG50 results, we suspect that these high NG50 and assembly length results are caused due to highly repetitive and inaccurate regions in these assemblies due to their poor genome fraction and average GC content results. The average GC content of the assemblies constructed using the \cc reads is significantly distant from the GC content of their corresponding reference genomes in most cases. This poor genome fraction and average GC content results \sr{suggest} that such large NG50 and assembly length values from \cc may also be caused \sr{by} poorly basecalled reads that lead to unresolved repetitive regions (i.e., bubbles in genome assembly graphs) or \sr{a} strong bias toward certain error types (i.e., homopolymer insertions of a certain base) in the assembly~\cite{ferrarini_evaluation_2013, chen_effects_2013}. \Copy{IR3.10/3}{\hgb{Fifth, the low  Total Indels and Indel Ratio (\%) for \mbox{\mech} in an assembled sequence signify a sequence that closely resembles the expected reference with minimal insertions and deletions (indels). This indicates a well-structured and high-quality assembly. Such assemblies  
offer a clear and accurate representation of the original sequence, facilitating downstream analyses, gene prediction, functional annotation, and comparative genomics.}} \Copy{IR3.7/2}{\hgb{Sixth, \mbox{\mech} consistently provides a higher quality value (QV), indicating a low probability of sequencing errors. Therefore, compared to the other evaluated basecallers, \mbox{\mech} has higher reliability of the assembled genome.}}


We conclude that, in most cases, the reads \sr{basecalled by} \mech lead to high\sr{er} quality, \sr{more} contiguous, and \sr{more} complete assemblies than that \sr{provided by other state-of-the-art basecallers,} \cc, \gpf, \bon, \sac, and \dor.

\subsubsection{Read mapping}
\label{subsubsection:down_read_mapping}
\hy{We provide the comparison of \mbox{\mech} with \cc, \gpf, \bon, \sac, and \mbox{\dor} in terms of the total number of base mismatches, the total number of mapped bases, the total number of mapped reads, and the total number of unmapped reads in Figure~\mbox{\ref{fig:samtool}}(a), \mbox{\ref{fig:samtool}}(b), \mbox{\ref{fig:samtool}}(c), and \mbox{\ref{fig:samtool}}(d), respectively.} \Copy{IR3.9/2}{\hgb{We also show the average read length, the overall number of mapped reads and the mapped bases, and the ratio of the number of mapped bases to the number of mapped reads in \fgb{Additional file 1:}  Table S2.}}

\begin{figure*}[h]
\setcounter{figure}{6}
\centering
\begin{subfigure}[h]{0.5\textwidth}
  \centering
  \includegraphics[width=\linewidth,trim={0.2cm 0.2cm 0cm 0.2cm},clip]{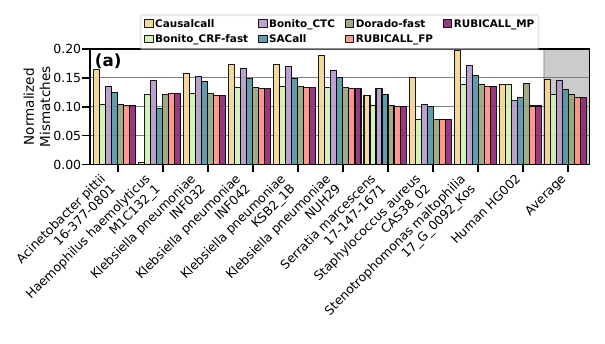}
  \vspace{-0.6cm}
\end{subfigure}%
\begin{subfigure}[h]{0.5\textwidth}
  \centering
  \vspace{-0.04cm}
  \includegraphics[width=\linewidth,trim={0.2cm 0.2cm 0.2cm 0.2cm}, clip]{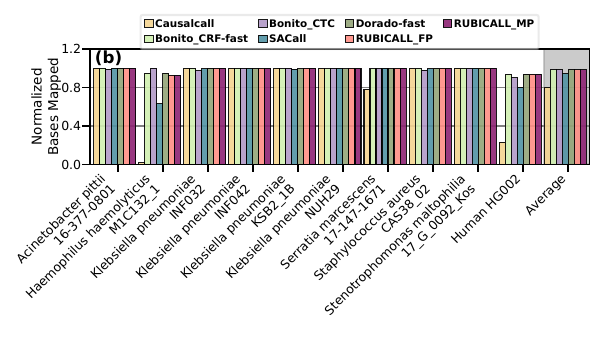}
  \vspace{-0.6cm}
\end{subfigure}

\begin{subfigure}[h]{0.49\textwidth}
  \centering
  \vspace{0.2cm}
  \includegraphics[width=\linewidth,trim={0.2cm 0.2cm 0.2cm 0.2cm}, clip]{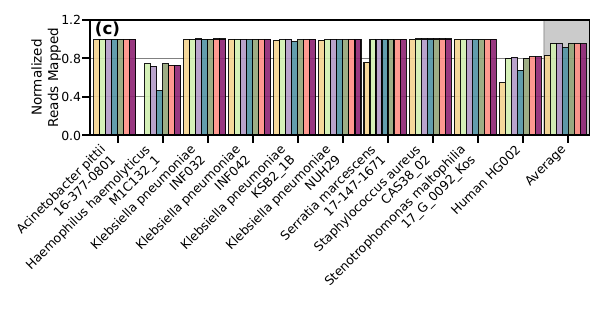} 
  \vspace{-0.6cm}
\end{subfigure}
\begin{subfigure}[h]{0.5\textwidth}
  \centering
  \vspace{-0.04cm}
  \includegraphics[width=\linewidth,trim={0.2cm 0.2cm 0.2cm 0.2cm}, clip]{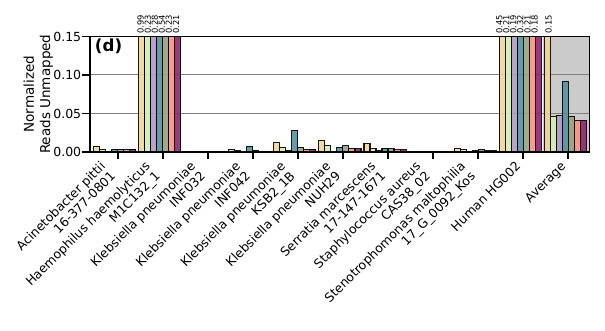}
  \vspace{-0.6cm}
\end{subfigure}
\caption{Comparison of \mech (using floating-point precision (\mechfp) and mixed-precision (\mechmp)) for normalized (a) mismatches, (b) bases mapped, (c) reads mapped, and (d) reads unmapped. 
\label{fig:samtool}}
\end{figure*}

We make five key observations. \hy{First, \mbox{\mech} provides the lowest number of base mismatches, which are 26.97\%, 22.66\%, 11.45\%, 12.35\%, and 23.58\% lower compared to \cc, \gpf, \bon, \sac, and \dor, respectively.} 
This indicates that \mech provides more accurate basecalled reads that share large similarity with the reference genome.
This is in line with the fact that \mech provides the highest basecalling accuracy, as we evaluate in Section~\ref{subsection:results_base_acc}.
\hy{Second, \mbox{\mech} provides, on average, 22.86\%,  0.24\%, and 4.77\% higher number of mapped bases compared to \cc, \bon,  and \sac, respectively, and only 0.3\% and 0.4\% lower number of mapped bases when compared to \gpf and \dor, respectively.}
Mapping more bases to the target reference genome confirms that the careful design and optimizations we perform when building \mech have no \hc{negative effects on} the basecalling accuracy.
\hy{Third, unlike \cc, \mech, \gpf, \bon, \sac, and \dor, all provide a high number of mapped reads.
However, \mbox{\mech} is the only basecaller that provides high-quality reads that have the highest number of base matches and the lowest number of base mismatches.}
\hy{Fourth, 
\mbox{\mech} achieves 72.66\%, 11.79\%, 14.63\%, 55.02\%, and 11.61\% lower unmapped reads compared to \cc, \gpf, \bon, \sac, and \mbox{\dor} respectively.
This indicates that using \cc, \gpf, \bon, \sac, and \mbox{\dor} wastes a valuable, expensive resource, i.e., sequencing data, by not mapping reads to the reference genome due to basecalling inaccuracies during basecalling.}
If a read is flagged as unmapped during read mapping, then this read is excluded from all the following analysis steps affecting the overall downstream analysis results.
\Copy{CC1/3}{\hgb{Fifth, for each dataset, we find that the ratio of the number of mapped bases to the number of mapped reads and the average length of the reads are mainly similar across all basecallers (\fgb{Additional file 1:}  Table S2), while \cc has a substantially lower ratio for the human genome. This mainly indicates that unaligned bases across basecallers are mainly shared within the mapped reads, resulting in a similar number of mapped reads with similar average lengths as well as the ratio.}}
We conclude that \mech reads provides the highest-quality read mapping results with the largest number of mapped bases and mapped reads.

\subsection{\strim analysis}
\begin{figure}[h]
  \centering
\includegraphics[width=\linewidth,trim={0.2cm 0.85cm 0.4cm 0.3cm},clip]{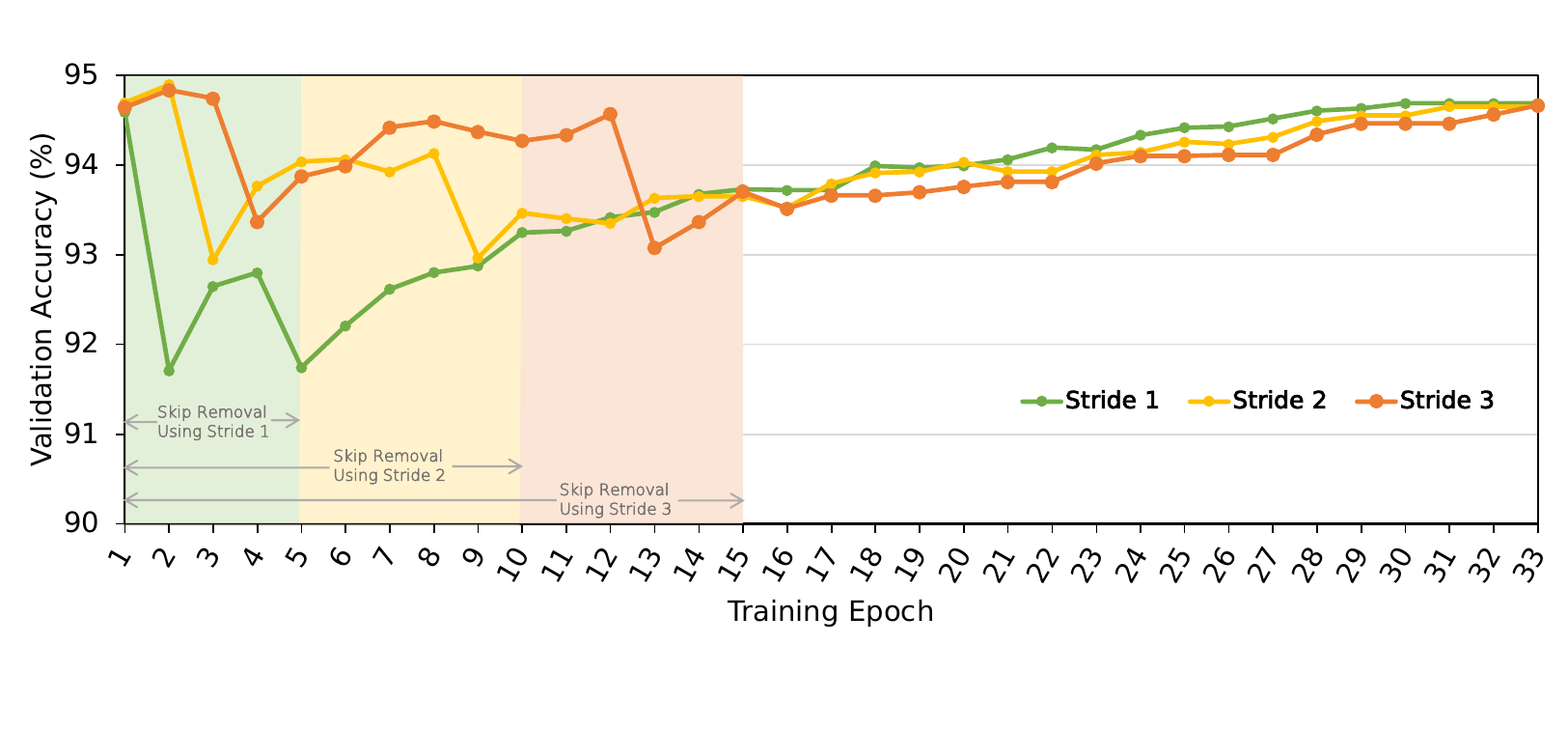}
  \vspace{-35pt}
  \caption{Effect of different strides while removing skip connections.}
  \label{fig:strim_analysis}
\end{figure}

\Copy{IR1.8/1}{Figure~\ref{fig:strim_analysis} shows the effect of \strim on validation accuracy using three different strides at which we remove a skip connection from a block, \hgb{i.e., the epoch interval at which \mbox{\strim} removes a skip connection from a block}.  We use our \nas-designed model that has five blocks of skip connections. \hgb{We highlight the number of epochs needed to remove all the skip connections for different strides. For example, \texttt{Stride 1} requires five epochs to remove all the skip connections, while \texttt{Stride 3} requires fifteen epochs.}} We make three observations. First, \texttt{Stride 1} converges faster to the baseline accuracy compared to \texttt{Stride 2} and \texttt{Stride 3}. By using \texttt{Stride 1}, we quickly remove all the skip connections (in five epochs) giving enough fine-tuning iterations for the model to recover its loss in accuracy. Second, all the strides show the maximum drop in accuracy (1.27\%-2.88\%) when removing skip connections from block 1 and block 4. We observe these blocks consist of the highest number of neural network model parameters due to the skip connections (30.73\% and 25.62\% of the total model parameters are present in skip connections in block 1 and block 4, respectively). Therefore, the model requires more training epochs to recover its accuracy after the removal of skip connections from these blocks. Third, a lower stride can get rid of skip connections faster than using a higher stride. However, all strides eventually converge to the baseline accuracy at the expense of more training iterations. We conclude that \strim provides an efficient mechanism to remove hardware-unfriendly skip connections without any loss in basecalling accuracy.

\subsection{Effect of pruning \mech}
\begin{figure}[h]
  \centering
  \includegraphics[width=1\linewidth,trim={0.2cm 0.85cm 0.4cm 0.3cm},clip]{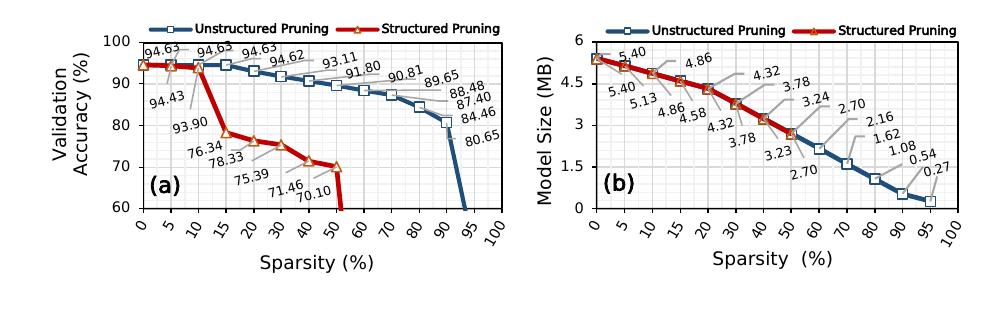}
  \caption{Effect of pruning \mech on: (a) validation accuracy and (b) model size.}
  \label{fig:res_prune_rubicon}
\end{figure}

Figure~\ref{fig:res_prune_rubicon} shows the effect of pruning \mech using two different pruning methods: unstructured element pruning  and structured channel pruning. 

We make four major observations. First, we can remove up to 15\% and 5\% of model parameters providing 1.18\% and 1.05\% reductions in model size without any loss in accuracy by using unstructured pruning and structured pruning, respectively. However,  unstructured pruning is unsuitable for hardware acceleration due to \hc{irregular structure}, and structured pruning provides minimal model size (or parameters) savings.  Therefore, we do not apply these pruning techniques to optimize \mech further. Second, we observe a drop in accuracy for pruning levels greater than 15\% and 5\% for unstructured and structured pruning, respectively. This shows that \nas found an optimal architecture as there is little room for pruning \mech further without loss in accuracy.

 Third, we observe that the \emph{knee point} for unstructured pruning  and structured pruning lies at 90\% and 50\%, where we achieve 80.65\% and 70.10\% of accuracy with 9.99$\times$ and 1.99$\times$ savings model size, respectively. After the knee point, we observe a sharp decline in accuracy. Fourth, below the knee point, we can trade accuracy for speed to further accelerate \mech for hardware computation and resources by removing unimportant network weights. We  conclude that pruning provides a tradeoff between accuracy and model size that can lead to further reductions in processing and memory demands for \mech, depending on the type of device on which genomic analyses would be performed. 

\subsection{Explainability into \nas results}
\begin{figure}[h]
  \centering
  \includegraphics[width=\linewidth,trim={0.2cm 0.85cm 0.4cm 0.3cm},clip]{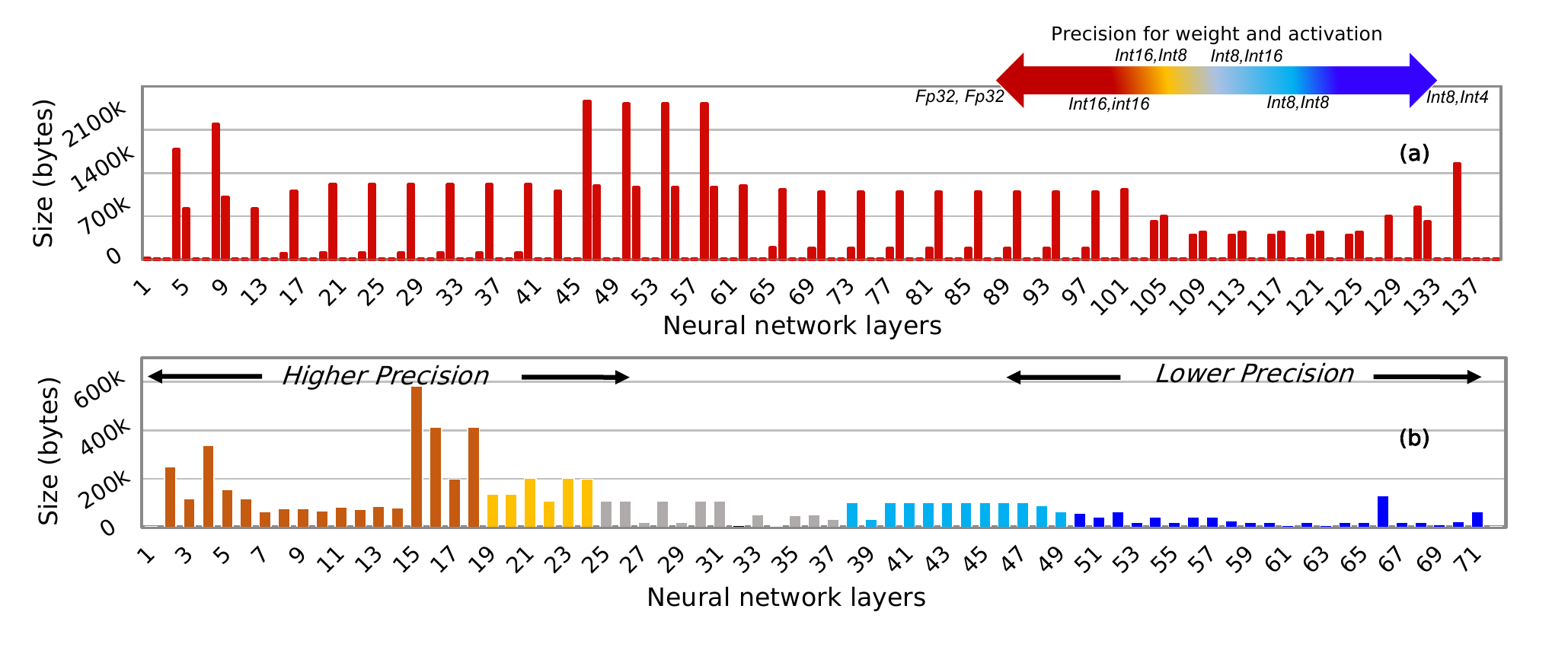}
  \caption{Layer size comparison for basecallers: (a) \bon, and (b) \mechmp.}
  \label{fig:explainability}
\end{figure}

We perform an explainability analysis to understand our results
further and explain \texttt{QABAS}’s decisions. 
The search performed by \nas provides insight into whether \nas has learned meaningful representations in basecalling. In Figure~\ref{fig:explainability}(a) and \ref{fig:explainability}(b), we extract the number of model parameters and precision of each parameter in a neural network layer to calculate the total size for each layer for \bon and \mechmp, respectively. \Copy{IR1.8/2}{\hgb{We highlight each layer's precision (i.e.,  weights and activation precision) using distinct colors. Our range includes floating-point (i.e., \texttt{fp32}) computation to integer computation (i.e., \texttt{int16}, \texttt{int8}, and \texttt{int4}) for weight and activation. Based on our experiments in  Section~\ref{suppsubsubsec:overprovision_quant}, we restrict the precision of weight and activation in \mechmp architecture in \nas to \texttt{int8} and \texttt{int4}, respectively.}} We compare \mechmp to \bon as it has the same backend (i.e, Quartznet~\cite{kriman2020quartznet}) and is designed by ONT experts. We make three observations. First, \nas uses more bits in the initial layers than the final layers in \mechmp. \nas learns that the input to \mech uses an analog squiggle that requires higher precision, while the output is only the nucleotide bases (A, C, G, T), which can be represented using lower precision. 

Second, \mech uses 1.97$\times$ less number of neural network layers than \bon while providing similar or higher basecalling accuracy on the evaluated species (Section~\ref{subsection:results_base_acc}). Thus, the superior performance of a basecaller architecture is not explicitly linked to its model complexity, and  \nas-designed models are parameter efficient.  Third, \bon uses the same single-precision floating-point representation (FP32) for all neural network  layers, which leads to very high memory bandwidth and processing demands. Whereas \mech has every layer quantized to a different quantization domain. We conclude that \nas provides an efficient automated method for designing more  efficient and hardware-friendly genomic basecallers compared to expert-designed basecallers.

\section{Discussion} \label{sec:discussion}
We are witnessing a tremendous transformation in 
high-throughput sequencing to significantly advance omics and  other life sciences. 
The bioinformatics community has developed a multitude of software tools to leverage increasingly large and complex sequencing datasets. Deep learning models have been especially powerful in modeling basecalling. 

\hy{\mbox{\head{Importance of basecalling}} Basecalling is the most fundamental computational step in the high-throughput sequencing pipeline. It is a critical problem in the field of genomics, and it has a significant impact on downstream analyses, such as variant calling and genome assembly. Improving the efficiency of basecalling has the potential to reduce the cost and time required for genomic analyses, which has practical implications for real-world applications. \mbox{\mech} offers a valuable alternative for researchers and practitioners who seek a balance between accuracy and speed. By maintaining competitive accuracy levels while significantly improving speed, our framework addresses the needs of various applications with stringent time constraints, ultimately benefiting a broader range of users. We believe that RUBICON provides a significant improvement over existing methods, and it has practical implications for the genomics community. 
}

\hy{\mbox{\head{Need to improve the throughput of basecallers}} Increasing throughput and reducing model size is critical because of the following three reasons. First, current basecallers already have high accuracy, but biologists do not pay attention to the throughput implications of using large deep learning-based models~\mbox{\cite{alser2022molecules}.}} We observe researchers building larger and larger  basecallers in an attempt to gain more accuracy without heeding to the disproportionately higher amount of power  these basecallers are consuming. Moreover,  none of the previous basecallers~\cite{wick2019performance,neumann2022rodan,konishi2021halcyon,xu2021fast,lou2020helix,perevsini2021nanopore,pages2022comprehensive,ambernas_zhang2021automated} have been optimized for mixed-precision execution to reduce energy consumption. As energy usage is proportional to the size of the network, energy-efficient basecalling is essential to enable the adoption of more and more sophisticated basecallers. \hy{Second, speed is critical in certain applications and use cases, particularly those that require real-time or near-real-time processing. \mbox{\framework} addresses these needs by focusing on hardware optimization and efficient implementation, ultimately enabling faster basecalling and potentially opening up new possibilities for applications with stringent time constraints. Third, as deep learning techniques and hardware continue to evolve, the balance between accuracy and speed/energy will remain an important aspect of model development. \mbox{\framework} provides a foundation for future research and innovation in hardware-friendly deep learning models for genomic basecalling.}

\hy{\mbox{\head{Evaluating \framework on other platforms}} All the state-of-the-art basecallers and RUBICON use high-level libraries, such as PyTorch or TensorFlow, which abstract the hardware architecture and provide a unified interface for deep learning computations.  These libraries work out-of-the-box for AMD GPUs and are equally optimized for them. Currently, high-level libraries do not provide capabilities to exploit low-precision tensor cores available on the latest GPUs. As a result, existing basecallers take advantage of comparable architectural capabilities regardless of the specific GPU employed.   Therefore, the hardware and software optimizations are at the same level for all supported GPU-based platforms. 
}

\hy{\mbox{\head{Automating basecaller generation process}}} Modern basecallers generally employ convolution neural networks to extract features from raw genomic sequences. However, designing a basecaller comes with a cost that a neural network model can have many different computational elements making the neural network tuning a major problem. 
At present, the vast majority of deep learning-based basecallers are manually tuned by computational biologists through manual trial and error, which is time-consuming. To a large extent, basecallers are being designed to provide higher accuracy without considering the compute demands of such networks. Such an approach leads to computationally complex basecallers that impose a substantial barrier to performing end-to-end time-sensitive genomic analyses.
This vast dependence of computational biologists \sr{and biomedical researchers} on these deep learning-based models creates a critical need to find efficient basecalling architectures optimized for performance. 

During our evaluation, we ran \nas for 96 GPU hours to sample architectures from our search space. Using complete sampling to evaluate all the 1.8$\times$10$^{32}$ viable options would take at least $\sim$4.3$\times$10$^{33}$ GPU hours. Thus, \nas accelerates the basecaller architecture search to develop high-performance basecalling architectures.   The final model architecture  can be further fine-tuned for other hyperparameters~\cite{singh2019napel,10.1145/3470496.3527442}, such as learning rate and batch size (for example, with grid search or neural architecture search). 
Throughout our experiments, we build general-purpose basecalling models by training and testing the model using an official, open-source ONT dataset that consists of a mix of different species. We did not specialize basecalling models for a specific specie. Past works, such as~\cite{wick2019performance}, show that higher basecalling accuracy can be achieved by building species-specific models.

\hy{\mbox{\head{Extending \framework}} \framework's modular design allows for the incorporation of additional layers or techniques, such as RNN, LSTM, and Transformers, to potentially increase accuracy further.}  \hy{We focus on convolution-based networks because: (a) matrix multiplication is the fundamental operation in such networks that is easily amenable to hardware acceleration, (b) the training and inference of RNN and LSTM models inherently involve sequential computation tasks, which poses a challenge for their acceleration on contemporary hardware such as GPUs and field-programmable gate arrays (FPGAs)~\mbox{\cite{nurvitadhi2016accelerating,singh_fpga-based_2021,singh2023sparta,singh2022accelerating,cali_segram_2022,singh2018review,singh2019near,gomez2023evaluating,singh2021modeling}}, and (c) transformer-based models are typically composed of multiple fully connected layers, which can be supported in \mbox{\framework} by modifying convolutional layers for improved computational efficiency and performance \mbox{\cite{umuroglu2017finn}}.} As future work,    \nas can be extended in two ways: (1) evaluate advance model architectures (such as RNN, transformer, etc.), and (2) perform more fine-grain quantization. First, extending \nas to other model architectures is important for researchers to quickly evaluate different computational elements. As the field of machine learning is rapidly evolving, it is non-trivial for researchers to adapt their models with the latest deep learning techniques. Second, currently, we perform mixed precision quantization, where every layer is quantized to a different domain. In the future, we can quantize every dimension of the weights to different precision. Such an approach would increase the design space of neural network architectural options to many folds.  \nas enables easy integration to explore such options automatically. Thus, \nas is easily extensible and alleviates the designer’s burden in exploring and finding sophisticated basecallers for different hardware configurations. We would explore two future directions for pruning a basecaller. First,  currently, we perform one-shot pruning, whereby we prune the model once and then fine-tune the model until convergence. Another approach could be to perform iterative pruning, where after every training epoch, we can re-prune the model using certain pruning criteria. Such an approach would further evaluate the fine-grained pruning limit of a basecaller. Second, an interesting future direction would be to combine multiple pruning techniques, e.g., structured channel pruning with structured group pruning (where we maintain the structure of the tensors without causing sparsity). Such an approach could lead to higher pruning ratios without substantial accuracy loss. 

\hy{\mbox{\head{Importance of \mech beyond basecalling}}} For \strim, we demonstrate its   applicability on  basecalling only, while there are other genome sequencing tasks where deep learning models with skip connections are actively being developed, such as predicting the effect of genetic variations~\cite{alipanahi2015predicting,ambernas_zhang2021automated}, detecting replication dynamics~\cite{boemo2021dnascent}, and predicting super-enhancers~\cite{sabba2021residual}. In \fgb{Additional file 1:} Section S1, we show the effect of manual skip removal, where we manually remove all the skip connections at once. We observe that the basecaller achieves 90.55\% accuracy (4.08\% lower than the baseline model with skip connections). By manual skip removal, the basecaller is unable to recover the loss in accuracy because CNN-based basecallers are sensitive to skip connections. Therefore, \strim provides a mechanism to develop hardware-friendly deep learning models for other genomic tasks.

\Copy{IR2.2}{\ogb{\head{Separation between \nas and \strim} Both \nas and \strim share the overarching objective of creating a compact basecalling network without compromising accuracy. However, they approach this goal from distinct perspectives and employ different optimization tools. The following three points justify the separation of the two methods. First, skip connections are integral to stable model training, and by retaining them during the initial \nas phase, we ensure effective training of the final basecalling network. The subsequent application of \strim allows for the controlled removal of skip connections, contributing to a more robust solution.  Second,  \nas might find an architecture with skip connections, whereas \strim employs knowledge distillation for skip connection removal, addressing a specific aspect not efficiently handled by \nas alone.  Third, unlike \strim, \nas tailors the neural network architecture for hardware efficiency without relying on a teacher network. The teacher network provides an upper bound on the achievable accuracy. Therefore, this two-step approach optimally combines the strengths of NAS and knowledge distillation, ensuring a comprehensive and effective optimization process for a compact and efficient basecalling model.}}



\section{Conclusion}

Nanopore sequencing generates noisy electrical signals that require conversion into a standard DNA nucleotide base string through a computational process known as basecalling. Efficient basecalling is crucial for subsequent genome analysis steps. Current basecalling approaches often neglect computational efficiency, resulting in slow, inefficient, and resource-intensive basecallers. To address this, we present \framework, a framework designed for creating hardware-optimized basecallers. \framework introduces two novel machine-learning techniques: \nas, an automatic architecture search for computation blocks and optimal bit-width precision, and \strim, a dynamic skip connection removal module that significantly reduces resource and storage requirements without sacrificing basecalling accuracy. We demonstrate the capabilities of \nas and \strim by designing
\mech, the first hardware-optimized basecaller, demonstrates fast, accurate, and efficient basecalling, achieving $\sim$6.88$\times$ reductions in model size with 2.94$\times$  fewer neural network parameters compared to an expert designed basecaller. We believe our open-source implementations of \framework will inspire advancements in genomics and omics research and development.
\section{Methods} \label{sec:evaluation}


\head{Evaluation setup}
Table~\ref{tab:systemparameters} provides our system details. We evaluate \mech using:  (1) AMD MI210 GPU~\cite{mi210} (\mechfp) \gssss{using floating-point precision computation},  and  (2) Versal ACAP VC2802~\cite{aiemlvc2802}, a  cutting-edge spatial vector computing system  from AMD-Xilinx (\mechmp) \gssss{using mixed-precision computation}. 
The 
Versal ACAP VC2802 features Versal AI Engine ML (AIE-ML)~\cite{aiemlvc2802} with 304 cores. The AIE-ML vector datapath implements two-dimensional single instruction, multiple data (SIMD)~\cite{illiac_simd_1968} operations using precisions ranging from int4$\times$int8 to int16$\times$int16 
operands that can execute 512 to 64 multiply-accumulate operations (MACs) per cycle, respectively. \gont{With its many different datatype precision options, AIE-ML acts as a suitable platform to demonstrate the benefits of a mixed precision basecaller.} We train all the basecallers \hy{(\cc, \gpf, \bon, \sac, and \dor)} using the same MI50 GPU.  \gssss{We use ONNX (Open Neural Network
 Exchange)~\cite{onnx} representation to evaluate the performance  on AIE-ML by calculating bit operations (BOPs)~\cite{baskin2021uniq}, which 
measures the number of bitwise operations in a given network, take into account the total number of supported operations per datatype on AIE-ML.}  
  
\vspace{0.6cm}
\begin{table}[htbp]
\caption{System parameters and hardware configuration for the CPU, GPU, and the AMD-Xilinx Versal ACAP.}
\vspace{-0.6cm}
    \label{tab:systemparameters}
      \begin{center}
    \renewcommand{\arraystretch}{1}
\setlength{\tabcolsep}{2pt}
    \resizebox{0.8\linewidth}{!}{%
\begin{tabular}{|c|c|}
    \hline
    \textbf{CPU} 
     & AMD EPYC 7742~\cite{amdEPYC} \\&@2.25GHz, 4-way SMT~\cite{smt}\\
     \hline
     \textbf{Cache-Hierarchy}&32$\times$32 KiB L1-I/D, 512 KiB L2, 256 MiB L3 \\
     \hline
     \textbf{System Memory}&4$\times$32GiB RDIMM DDR4 2666 MHz~\cite{rdimm} PCIe 4.0 $\times$128 \\
    \hline
     \textbf{OS details}& 
      \begin{tabular}[c]{@{}l@{}}
         Ubuntu 21.04 Hirsute Hippo~\cite{ubuntu},\\ GNU Compiler Collection (GCC) version 10.3.0~\cite{gnu}
     \end{tabular}  \\
      \hline
      \hline
     \textbf{GPU} &    \begin{tabular}[c]{@{}l@{}}
     AMD Radeon Instinct™ MI210~\cite{mi210} 6656 Stream Processors@1.7GHz\\ 64GB HBM2 PCIe 4.0 $\times$16, ROCm version 5.1.1~\cite{rocm511} \\ \hline
     NVIDIA A40~\cite{a40} 10,752 CUDA Cores@1.2GHz, 48GiB DRAM\\
     NVIDIA System Management Interface (NVIDIA-SMI) version 510.47.03~\cite{nvidia-smi}\\
     NVIDIA CUDA Compiler Driver (NVCC) version 11.4~\cite{nvcc}
          
         
     \end{tabular}  \\ \hline
     \hline
     \textbf{AMD-Xilinx Versal ACAP} &    \begin{tabular}[c]{@{}l@{}}
         Versal ACAP VC2802~\cite{aiemlvc2802}, 304$\times$AIE-ML@1GHz,\\
         19MB local memory, Dual-Core Arm Cortex-A72\cite{armOnAIE}
     \end{tabular}  \\ \hline
\end{tabular}
}
  \end{center}
 \end{table}

\head{\nas setup details
} We use the publicly available ONT dataset~\cite{bonito} 
\sr{sequenced using} MinION Flow Cell (R9.4.1) for the training and validation during the \nas search phase. \Copy{IR3.11}{\ogb{The dataset comprises 1,221,470 reads, all sequenced from complete genomes. This ONT training dataset has an approximate list of 496 unique taxonomic IDs using the Kraken2~\cite{kraken2} taxonomic classification system~\cite{palamut_2020}.}} We randomly select 30k samples from the training set for the search phase (specified using the \texttt{---chunks} parameter). We use nni~\cite{nni} with nn-meter~\cite{nnmetercode} to implement hardware-aware NAS. We use the Brevitas library~\cite{brevitas} to perform quantization-aware training.
 The architectural parameters and network weights are updated using AdamW~\cite{kingma2014adam} optimizer with  a learning rate of  2$e^{-3}$, a beta value of 0.999, \sr{a} weight decay of 0.01, \sr{and an} epsilon of 1$e^{-8}$. We set the hyperparameter $\lambda$ to 0.6. \gs{We choose these values based on our empirical analysis.} 
After the \nas search phase, the sampled networks are trained \sr{until convergence with knowledge distillation} using the same ONT dataset that we use during the \nas search phase, 
with a batch size \sr{of} 64, \gs{based on the maximum memory capacity of our evaluated {Mi50 GPU}}.  We set knowledge distillation hyperparameters alpha ($\alpha$) and temperature ($\tau$) at 0.9 and 2, respectively. 

 \head{\nas search space} For the computations operations, we search for a design with one-dimensional (1D) convolution with ten different options: kernel size (KS) options (3, 5, 7, 9, 25, 31, 55, 75, 115, and 123) for grouped 1-D convolutions. We also use an identity operator that, in effect, removes a layer to get a shallower network. For quantization bits, we use  bit-widths that are a factor of 2$^{n}$, where 2<n<4 (since we need at least 2 bits to represent nucleotides A, C, G, T and 1 additional bit to represent an undefined character in case of a misprediction). We use four different quantization options for weights and activations ($<$\texttt{8,4}$>$, $<$\texttt{8,8}$>$, $<$\texttt{16,8}$>$, and $<$\texttt{16,16}$>$). We choose these quantization levels based on the precision support provided by our evaluated hardware and the effect of quantization on basecalling (Section~\ref{supfig:over_quant}). We use five different channel sizes with four repeats each. We choose the number of repeats based on the maximum memory capacity of our evaluated GPU. In total, we have $\sim$1.8$\times$10$^{32}$ distinct model options in our search space $\mathcal{M}$. 
 
\head{\strim details}\gsss{ We use \bon as the teacher network, while the \nas-designed model is the student network. We remove skip connections with a stride 1 (using parameter \texttt{---skip\_stride}). Based on hyper-parameter tuning experiments (\fgb{Additional file 1:} Section S2), set knowledge distillation hyperparameters alpha ($\alpha$) and temperature ($\tau$) at 0.9 and 2, respectively. We use Kullback-Leibler divergence loss to calculate the loss~\cite{kldivloss}}.

\head{Pruning details} We use PyTorch~\cite{pytorch} modules for both unstructured and structured pruning~\cite{torch_modules} with L1-norm, i.e., prune the weights that have the smallest absolute values. We apply one-shot pruning, where we first prune a model with a specific amount of sparsity, then train the model until convergence on the full ONT dataset~\cite{bonito}.

\head{Baseline basecallers} \hy{\mbox{\mech} is a pure convolution-based network. We focus on convolution-based networks because: (a) matrix multiplication is the fundamental operation in such networks that is easily amenable to hardware acceleration, (b) the training and inference of RNN and LSTM models inherently involve sequential computation tasks, which poses a challenge for their acceleration on contemporary hardware such as GPUs and field-programmable gate arrays (FPGAs)~\mbox{\cite{nurvitadhi2016accelerating}}, and (c) transformer-based models are typically composed of multiple fully connected layers, which can be supported in \mbox{\framework} by modifying convolutional layers for improved computational efficiency and performance \mbox{\cite{umuroglu2017finn}}.}  We compare \mech against five different basecallers: (1) \cc~\cite{zeng2020causalcall} is a state-of-the-art  basecaller with skip connections, (2) \gpf~\cite{bonito} v0.6.2 is a 
recurrent neural network (RNN)-based version of basecaller from ONT that is optimized for throughput  for real-time basecalling on Nanopore devices,  (3) \bon~\cite{bonito}  v0.6.2  is 
convolutional neural network (CNN)-based hand-tuned basecaller from ONT, \hy{(4) \mbox{\sac~\cite{huang2020sacall}} is a transformer-based basecaller that uses an attention mechanism for basecalling, and (5) 
 \dor\mbox{\cite{dorado}}  v0.4.0  is a LibTorch~\mbox{\cite{libtorch}} version of \mbox{\gpf} from ONT. \mbox{\dor} uses the same model architecture as \mbox{\gpf} and uses the \mbox{\texttt{Bonito}} framework for model training.} \cc and \bon uses the same backend structure as \mech (i.e., Quartznet~\cite{kriman2020quartznet}).  We are aware of other basecallers such as \texttt{Halcyon}~\cite{konishi2021halcyon},  
\texttt{Helix}~\cite{lou2020helix}, and \texttt{Fast-bonito}~\cite{xu2021fast}
. However, these basecallers are either not open-source or  do not provide training code with support for specific read formats. 

\head{Basecalling reads} 
To evaluate basecalling performance, we use a set of reads generated using a MinION R9.4.1 flowcell. \Copy{IR2.4}{\ogb{We use only R9 chemistry datasets as, currently, ONT does not provide a suitable public training dataset for R10 chemistry. They offer in-house trained R10 models that cannot be employed for a consistent evaluation across all basecallers. R9 and R10 chemistries involve distinct generations of nanopore technologies, including different pore proteins and read lengths. Therefore, models trained on R9 chemistry are incompatible for inference on R10 sequenced datasets. Due to these technical constraints, our study is currently limited to utilizing the available R9 chemistry training dataset from ONT and conducting inference exclusively on R9 chemistry datasets.}} Table~\ref{tab:basecall_reads} provides details on different organisms used in our evaluation. \hgb{We use several bacterial species and the human genome. For Human HG002, we use 3$\times$ depth of coverage.}  

\vspace{10pt}
\begin{table}[tbh]
\centering
 \renewcommand{\arraystretch}{0.45}
\setlength{\tabcolsep}{1.5pt}
 \setstretch{0.8}
\caption{Details of datasets used in evaluation.} \label{tab:basecall_reads}
\resizebox{0.6\columnwidth}{!}{
\begin{tabular}{@{}llll@{}}\toprule
\textbf{Organism} & \textbf{Chemistry} & \textbf{\# Reads}  & \textbf{Reference}\\
\textbf{} & \textbf{} & \textbf{} & \textbf{Genome} \\
\textbf{} & \textbf{} & \textbf{} & \textbf{Size (bp)} \\\midrule
\begin{tabular}{l}
  Acinetobacter pittii\\
  16-377-0801 
\end{tabular}  & R9.4.1 & 4,467   & 3,814,719 \\\cmidrule{1-4}
\begin{tabular}{l}
  Haemophilus haemolyticus\\
  M1C132\_1
\end{tabular}  & R9.4 & 8,669   & 2,042,591 \\\cmidrule{1-4}
\begin{tabular}{l}
  Klebsiella pneumoniae\\
   INF032 
\end{tabular} & R9.4 & 15,154   & 5,111,537 \\\cmidrule{1-4}
\begin{tabular}{l}
  Klebsiella pneumoniae\\
   INF042 
\end{tabular}   & R9.4 & 11,278   & 5,337,491 \\\cmidrule{1-4}
\begin{tabular}{l}
  Klebsiella pneumoniae\\
  KSB2\_1B
\end{tabular} & R9.4 & 15,178   & 5,228,889 \\\cmidrule{1-4}
\begin{tabular}{l}
 Klebsiella pneumoniae\\
 NUH29
\end{tabular} & R9.4 & 11,047   & 5,134,281 \\\cmidrule{1-4}
\begin{tabular}{l}
  Serratia marcescens\\
  17-147-1671
\end{tabular} & R9.4.1 & 	16,847   & 5,517,578 \\\cmidrule{1-4}
\begin{tabular}{l}
  Staphylococcus aureus\\
  CAS38\_02
\end{tabular} & R9.4.1 & 	16,742	  & 2,902,076 \\\cmidrule{1-4}
\begin{tabular}{l}
  Stenotrophomonas maltophilia\\
  17\_G\_0092\_Kos 
\end{tabular} & R9.4 & 16,010   & 4,802,733 \\\cmidrule{1-4}

\begin{tabular}{l}
  Human\\
  HG002 
\end{tabular} & R9.4.1 & 300,000   & 2,947,743,500 \\
\bottomrule
\end{tabular}}
\end{table}

\Copy{IR1.3/1}{\hgb{Prior to basecalling, raw nanopore signals undergo a preprocessing pipeline to prepare them for input into the neural network. Raw nanopore signals, which can be hundreds of thousands of data points long, are normalized to ensure consistent input characteristics for the subsequent processing steps. We use empirically determined normalization scaling factors from ONT's \bon basecaller. The normalized signals are chunked into smaller segments, typically with overlapping regions.}} \Copy{IR1.3/3}{\hgb{The chunk size and overlap are empirically set to 4000 bps and 500, respectively. Chunk size affects the balance between processing speed and accuracy. Smaller chunk sizes can lead to more accurate basecalling but may require more computational resources and time. Larger chunk sizes may be faster but can potentially introduce errors if the signal varies significantly within the chunk. Overlap represents the degree to which consecutive chunks share data with each other. Overlapping chunks can help mitigate the potential issues caused by abrupt changes in the signal at chunk boundaries. It allows for a smoother transition between chunks, reducing the chances of missing important information in the signal. However, a larger overlap may increase computational demands and processing time.}} \Copy{IR1.3/2}{\hgb{After basecalling, the basecalled sequences obtained from individual signal segments are stitched back together to reconstruct the entire nucleotide sequence. The stitched sequences are then decoded to obtain the final basecalled sequences. We use the beam-search decoding\mbox{\cite{silvestre2021pair}} method to obtain the final basecalled sequences from stitched segments.}}

\head{Basecaller evaluation metrics} We evaluate the performance of \mech
using two different metrics: 
(1) basecalling throughput (kbp/sec), i.e., the throughput of a basecaller in terms of kilo basepairs generated per second, and (2) basecalling accuracy (\%), i.e., the total number of bases of a read that are exactly matched to the bases of the reference genome divided by the total length of its alignment including insertions and deletions.
We measure the basecalling throughput for the end-to-end basecalling calculations, including reading FAST5 files and writing out FASTQ or FASTA file using Linux \textit{/usr/bin/time -v} command.
For basecalling accuracy, we align each basecalled read to its corresponding reference genome of the same species using the state-of-the-art read mapper, minimap2~\cite{li_minimap2_2018}. We use Rebaler~\cite{rebaler} to generate a consensus sequence from each basecalled read set, which replaces portions of the reference genome with read-derived sequences. The assembled genome is then polished with multiple rounds of Racon~\cite{vaser2017fast}. This results in an assembled genome that accurately represents the original data while minimizing potential errors introduced by the reference. 

\head{Downstream analysis} 
We evaluate the effect of using \mech and other baseline basecallers on two widely-used downstream analyses, \emph{de novo} assembly~\cite{robertson2010novo} and read mapping~\cite{li2010rna}.

\textbf{\emph{De novo} assembly.}
We construct \emph{de novo} assemblies from the basecalled reads and calculate the statistics related to the accuracy, completeness, and contiguity of these assemblies. 
To generate \emph{de novo} assemblies, we use minimap2~\cite{li_minimap2_2018} to report all read overlaps and miniasm~\cite{li_minimap_2016} to construct the assembly from these overlaps. 
We use miniasm because it allows us to observe the effect of the reads on the assemblies without performing additional error correction \sr{steps} on \sr{input} reads~\cite{firtina_hercules_2018} and the\sr{ir final} assembly~\cite{firtina_apollo_2020}. 
\sr{To measure the assembly accuracy, we use dnadiff~\cite{marcais_mummer4_2018} to evaluate} 1)~the portion of the reference genome that can align to a given assembly (i.e., Genome Fraction), 2)~the average identity of assemblies (i.e., Average Identity) when compared to their respective reference genomes, and \Copy{IR3.10/1}{\hgb{3) insertions and deletions of nucleotides (or bases) in the sequence when compared to a reference or other sequences. ( i.e., Total Indels and Indel Ratio (\%)). Total Indels represents the sum of all the insertions and deletions in the assembled sequence when compared to a reference or other sequences. The Indel Ratio is a measure of the relative abundance of indels compared to the total length of the assembled sequence (calculated using Total Indels / Assembly Length) $\times$ 100. This metric helps to understand the proportion of the assembly that contains insertions and deletions.}} 
To measure statistics related to the contiguity and completeness of the assemblies, such as the overall assembly length, average GC content (i.e., the ratio of G and C bases in an assembly), and NG50 statistics (i.e., shortest contig at the half of the overall reference genome length), we use QUAST~\cite{gurevich_quast_2013}. 
We assume that the reference genomes are high\sr{-}quality representative of the sequenced samples that we basecall the reads from when comparing assemblies to their corresponding reference genomes. \sr{The higher the values of the average identity, genome fraction, and NG50 results, the higher the quality of the assembly and, hence the better the corresponding basecaller.}
When the values of the average GC and assembly length results are closer to that of the corresponding reference genome, the better the assembly and the corresponding basecaller. \Copy{IR3.7/1}{\hgb{We use Inspector\mbox{\cite{chen2021accurate}} to calculate the overall quality value (QV) of an assembly. The QV score is determined by considering structural and small-scale errors in proportion to the total number of base pairs in the assemblies. High-quality sequences have higher QV scores, indicating a low probability of sequencing errors, while low-quality sequences have lower QV scores, suggesting a higher likelihood of errors.}}

\textbf{Read mapping.}
We basecall the raw electrical signals into reads using each of the subject basecallers. 
We map the resulting read set to the reference genome of the same species using the state-of-the-art read mapper, minimap2~\cite{li_minimap2_2018}. 
We use the default parameter values for mapping ONT reads using the preset parameter \textit{-x map-ont}.
We use the \textit{stats} tool from the SAMtools library~\cite{li2009sequence} to obtain four key statistics on the quality of read mapping results, the total number of mismatches, the total number of mapped bases, the total number of mapped reads, and the total number of unmapped reads. \Copy{CC1/2}{\hgb{We normalize the total number of base mismatches and the total number of mapped bases using the total number of bases in the reads, while for the total number of mapped reads and the total number of unmapped reads, we normalize using the total number of reads.}} 



\section{Availability of data and materials}
\label{sec:data_avail}
The read set and reference set used in this study \fgb{are part of work carried out by Wick et al.~\cite{wick2019performance},} which can be downloaded from \url{https://bridges.monash.edu/articles/dataset/Raw_fast5s/7676174} and \url{https://bridges.monash.edu/articles/dataset/Reference_genomes/7676135}, respectively. 
\Copy{CC1/1}{\hgb{For the human genome~\cite{rhie2023complete}, we download reads from \mbox{\url{https://labs.epi2me.io/gm24385_2020.11/}}, while the reference genome is available at \mbox{\url{https://github.com/marbl/HG002}}.}} All trained models and generated reads can be downloaded from \url{https://zenodo.org/record/10198815}.
We ensure unbiased, fair, and consistent evaluation by retraining all the basecallers using the official ONT dataset~\mbox{\cite{bonito}}.

\hgb{Source code with the instructions for reproducing the results is publicly available at: 
 \fgb{GitHub~\cite{rubiconCode} and Zenodo~\cite{zenadoCode}}}.
Scripts used to perform basecalling accuracy analysis are available at: \url{https://github.com/rrwick/Basecalling-comparison}.

\section{Acknowledgments}
\gsssss{We thank the SAFARI Research Group members for their valuable feedback and the stimulating intellectual \gssssss{and} scholarly environment they provide. SAFARI Research Group acknowledges the generous gifts of their industrial partners, including Google, Huawei, Intel, Microsoft, VMware, and AMD. This research was partially supported by the Semiconductor Research Corporation. SAFARI Research Group acknowledges support from the European Union’s Horizon programme for research and innovation under grant agreement No. 101047160, project BioPIM.}
 Special thanks to  Alessandro Pappalardo for his support with quantization-aware training. We appreciate valuable discussions with Giovanni Mariani.  Thanks to AMD for providing access to the HPC fund cluster~\cite{hpcfund}. 

\section{Author contributions}
 G.S., M.A., K.D., and A.K.  conceived \framework. G.S. designed and implemented \framework. 
 G.S., C.F., and M.C. collected data and performed the evaluations.  K.D., H.C., and O.M. supervised the work. 
 G.S., M.A.,  K.D., and C.F. wrote the manuscript. 
All authors reviewed and edited the manuscript. All authors analyzed the results. All authors read and approved the final manuscript.  

\fgb{
\section{Ethics approval and consent to participate}
Not applicable - ethical approval was not needed for the study, as publicly available datasets were used. No private, confidential, or sensitive information pertaining to individuals was utilized. Furthermore, our research did not involve any animal experiments.
}
\section{Competing interests} 
\fgb{Gagandeep Singh, Kristof Denolf, and Alireza Khodamoradi are affiliated with AMD.}
The remaining authors declare no competing interests.

\fgb{
\section{Supplementary information}
Additional 1: Supplementary notes S1-S6, Figures S1-S9, and Tables S1-S3.
}


\newpage
\balance
\bibliographystyle{vancouver}

{\small \bibliography{ref}}
\onecolumn
\setcounter{secnumdepth}{3}
\clearpage
\begin{center}
\textbf{\LARGE Supplementary Material for\\ \ltitle}
\end{center}
\setcounter{section}{0}
\setcounter{equation}{0}
\setcounter{figure}{0}
\setcounter{table}{0}
\setcounter{page}{1}
\makeatletter
\renewcommand{\theequation}{S\arabic{equation}}
\renewcommand{\thetable}{S\arabic{table}}
\renewcommand{\thefigure}{S\arabic{figure}}
\renewcommand{\thesection}{S\arabic{section}}
\renewcommand{\thesubsection}{S\arabic{subsection}}
\renewcommand{\thesubsubsection}{S\arabic{subsubsection}}

\newcommand{\TextUnderscore}{\rule{.4em}{.4pt}}

\section{Quantization-aware basecaller architecture search (\nas)}
\label{supsec:qabas}
\begin{figure*}[h]
  \centering
  \includegraphics[width=\linewidth]{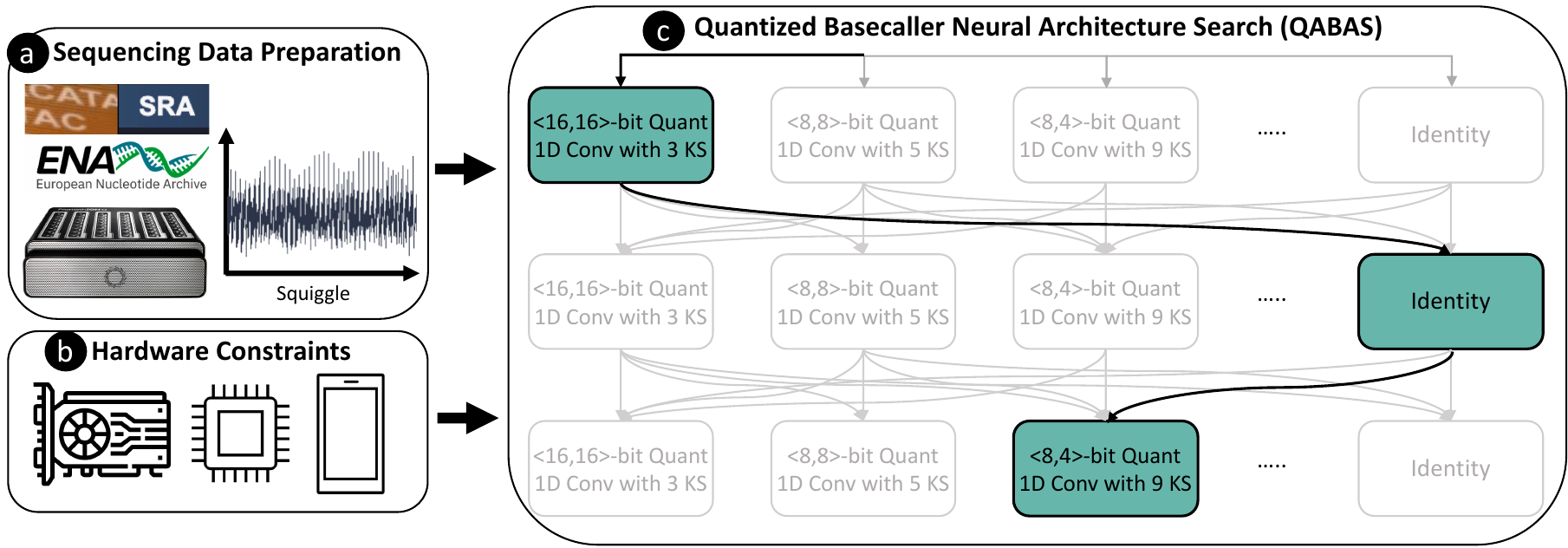}
  \vspace{-10pt}
  \caption{Overview of \nas. \nas evaluates a different set of candidate operations for convolution (\texttt{conv}) and quantization bits. In the figure, we show different options for kernel size (\texttt{KS}) (e.g., 3, 5, 9, etc.) and quantization bits  (4-b, 8-b, and 16-b) for each network layer. The identity operator removes a layer to get a shallower network. }
  \label{fig:nas}
  \vspace{12pt}
\end{figure*}

\nas automates the process of finding efficient and high-performance hardware-aware genomics basecallers.
\fgb{Additional file 1:} Figure~\ref{fig:nas} shows the workflow overview of \nas. The raw sequencing data \circled{a} is provided as input to \nas, which can be obtained through sequencing a new sample, downloading from publicly-available databases, or computer simulation. \nas uses such a set of data as training ($\mathbb{D}_{train}$) and evaluation set ($\mathbb{D}_{eval}$) while automatically designing a basecaller. To achieve a basecaller design that provides high throughput, we add hardware constraints \circled{b}, in terms of latency or throughput, to \nas. 
 A hardware-aware basecaller can better use the underlying hardware features and greatly accelerate inference speed. As a result, it improves the overall basecalling efficiency.   

\nas \circled{c} leverages automated machine learning (AutoML) algorithms~\cite{zoph2016neural} using neural architecture search (NAS) to design an efficient hardware basecaller by exploring and evaluating different neural network architectures from a pre-defined search space. The search space $\mathcal{M}$ consists of the possible neural network architectural options while $\mathbb{M}$ $\in$ $\mathcal{M}$  is a sub-architecture from $\mathbb{M}$. The goal is to find an optimal sub-architecture $\mathbb{M}^\ast$ using Equation~\ref{eq:nas_train}  that minimizes the training loss ($\mathcal{L}_{train}$) while going over $\mathbb{D}_{train}$ and gives maximum accuracy with the  $\mathbb{D}_{eval}$.
\begin{equation}
 \mathbb{M}^\ast=\argmax_{\mathbb{M} \in \mathcal{M}}  Eval (\mathbb{M}, \argmin_{w^\ast} \mathcal{L}_{train} (w^\ast(\mathbb{M}), \mathbb{D}_{train}); \mathbb{D}_{eval})\label{eq:nas_train}
\end{equation}

where $w^\ast(\mathbb{M})$ represent the weights of sub-architecture $\mathbb{M}^\ast$.


\head{\nas search space} We define the search space $\mathcal{M}$ as sufficiently large to enable a powerful neural architecture search.  A larger space enables the search algorithm to cover more architectures to increase the chance of finding a powerful architecture. However, a larger search space makes converging more difficult for the search algorithm. 

Our model search space has sequentially connected blocks, where each block receives input from its direct previous block.  We formulate the NAS problem for hardware-aware genomics basecaller as finding: (a) the computational operations in each basic block\footnote{Our basic block consists of one-dimensional (1-D) convolution, batch normalization~\cite{ioffe2015batch}, and rectified linear unit (ReLU)~\cite{agarap2018deep}.} of a basecaller, including operations in a skip connection block, and (b) 
quantization bit-width for weights and activations for each neural network layer to perform low-precision computation. Quantization is the reduction of the bit-width precision at which calculations are performed in a neural network to reduce memory and computational complexity. Adding quantization exploration dramatically increases the model search space ($\sim$6.72$\times$10$^{20}$ additional viable options in our search space). However, performing a joint search for computational blocks and quantization bits is crucial because: (1) optimizing these two components in separate stages could lead to sub-optimal results as the best network architecture for the full-precision model is not necessarily the optimal one after quantization, and (2) independent exploration would also require considerable search time and energy consumption because of many viable design options~\cite{ren2021comprehensive}.  Therefore, \nas searches for both the computational operations present in each basic block of a basecaller and the quantization bits used by these computational operations.   \Copy{IR3.12/1}{\ogb{In doing so, we tailor the neural network architecture and computation to align with the hardware's capabilities.}}



\head{\nas search algorithm}  \nas evaluates different neural network architectures using differentiable neural architecture search (DNAS)~\cite{liu2018darts,luo2018neural,xie2021weight}. DNAS follows a weight-sharing approach of reusing weights of previously optimized architectures from the neural architecture search space. For example, if sub-architecture $\mathbb{M}_1$ has only one additional layer compared to sub-architecture $\mathbb{M}_2$. In such a scenario, $\mathbb{M}_1$  can use most weights from $\mathbb{M}_2$. Therefore, the search procedure gets accelerated in DNAS compared to training each sub-architecture individually.

DNAS formulates the entire search space as a super-network and distills a target network from this super-network. Traditional NAS approaches~\cite{zoph2016neural} often sample many different architectures from the search space and train each architecture from scratch to validate its performance. Such an approach requires heavy computational resources that could lead to thousands of GPU hours of overhead. One way to overcome this issue is to use NAS with heuristic-based methods~\cite{real2017large,xie2017genetic}, such as genetic algorithms that select individual architectures from the current \emph{population} to be \emph{parents} and uses them to produce the \emph{children} for the next generation. However, such methods still suffer from the problem of retraining each sample architecture from scratch.  Therefore, DNAS provides an efficient solution by sharing computation among different architectures, as many of them have similar properties.

In \nas, we construct an over-parameterized super-network with all possible candidate options. The super-network shares weights among sub-architecture.  During the search phase, \nas searches for the optimal: (a) architectural parameter $\alpha$: likelihood that a computational operation will be preserved in the final architecture; and (b) network weights $w$: weights of convolution layers. We use ProxylessNAS~\cite{cai2018proxylessnas}  to binarize the architectural parameter (i.e., $\alpha \in$ \{0,1\}) to reduce  memory consumption during the search phase. At the end of the search phase, the operators with the highest architectural weight are preserved, while others are eliminated. Since the NAS search procedure is focused on optimizing the super-network, the final sub-network architecture $\mathbb{M}^\ast$, with all the preserved  operations, is retrained to convergence to fully optimize its network weights. 

\head{Quantization-aware hardware metric} Current state-of-the-art basecallers~\cite{wick2019performance,neumann2022rodan,konishi2021halcyon,xu2021fast,lou2020helix,perevsini2021nanopore,pages2022comprehensive,ambernas_zhang2021automated} are hardware-agnostic. They only focus on improving the accuracy without paying attention to its inference efficiency. For example, Fast-bonito~\cite{xu2021fast} uses NAS for basecalling architecture search, however, it does not consider any hardware-related metrics during the architecture search. Therefore, such approaches lead to over-provisioned basecallers with a large number of parameters and model sizes that are unoptimized for mixed-precision computation (see Section~\ref{suppsec:overprovision}).  We overcome this inefficiency in \nas by adding hardware constraints, in terms of inference latency, to the \nas search phase. Thus, \nas aims to find an efficient neural network architecture for basecalling that is also optimized for hardware implementation. During the search process, \nas sequentially selects a sub-network from the super-network. The expected latency of the sub-network is the sum of the latencies of each operation in the network. Before the start of the \nas search phase, we profile the latencies of operations present in the search space on targeted hardware to build a latency estimator. We also incorporate the latency while using different quantization bit-widths for the weights and activations in our latency estimator.  This latency estimator is utilized to guide the \nas search process.  

\nas's objective function ($\mathcal{L}_{\nas}$) minimizes a joint cross-entropy error to: (a) provide better basecalling accuracy by minimizing the training loss ($\mathcal{L}_{train}$) while going over $\mathbb{D}_{train}$, and (b) minimize a regularization term ($\mathcal{L}_{reg}$) to find a sub-network $\mathbb{M}$ with inference latency ($\mathbb{L}_{\mathbb{M}}$) that satisfies our inference latency constraints.  We add  latency constraints by using a target latency parameter ($\mathbb{L}_{tar}$) to the regularization term $\mathcal{L}_{reg}$ to guide the search process. For example, in case we want a small model, then we can provide a higher $\mathbb{L}_{tar}$ value, or vice versa.



\begin{align*}
\mathcal{L}_{\nas}=\mathcal{L}_{train}+\lambda\mathcal{L}_{reg}\\
\mathcal{L}_{reg}=(\mathbb{L}_{\mathbb{M}}-\mathbb{L}_{tar})/\mathbb{L}_{tar}
\end{align*}

\noindent where $\lambda$ is a parameter to control the tradeoff between the basecalling accuracy and the model latency. 
\Copy{IR3.12}{\ogb{As different hardware provides different latencies for the same layers chosen from the \nas search space, the user can customize the \framework framework for their target hardware by adjusting hardware-specific parameters  (i.e., using the \texttt{applied\_hardware} flag in \framework~\cite{rubiconCode}). We provide an additional \texttt{reference\_latency} flag in \nas to guide the search of basecalling architecture to find an architecture that meets certain latency constraints.  This coupling of target hardware latency ensures the basecaller architecture is finely tuned to operate optimally on the intended hardware.}} \Copy{IR2.3}{\ogb{We provide an example latency estimator for our target hardware (i.e., AIE) in \framework. However, our integration with the open-source nn-Meter~\cite{nnmetercode} tool allows users to freely configure hardware settings through the \texttt{applied\_hardware} flag in \framework. This integration enhances adaptability, enabling efficient optimization and deployment across different hardware environments.}}

\section{\strim: Skip connection removal by teaching}
\label{supsec:skip}

\begin{figure*}[h]
  \centering
  \includegraphics[width=0.9\linewidth]{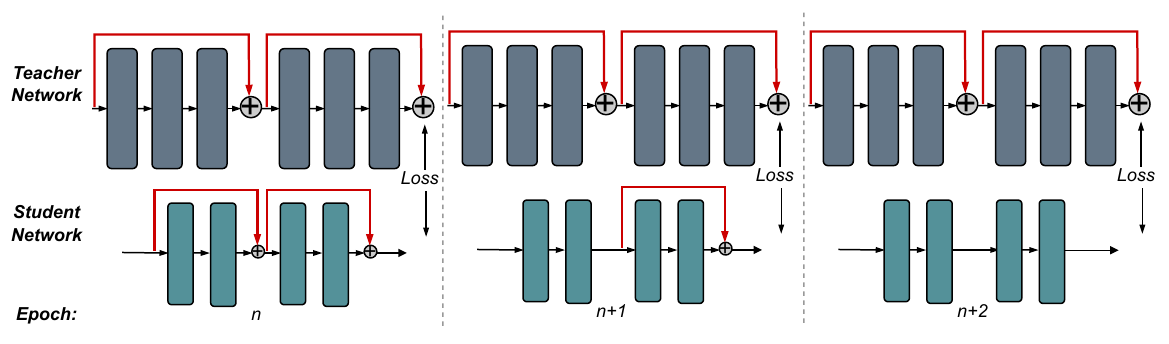}
  \caption{Overview of \strim process for three epochs. We start with a large, overprovisioned floating-point precision model as the teacher network and our \nas mixed-precision model as the student network. During the training, \strim removes a skip connection from the student network every \texttt{n} epoch, starting with the first skip connection encountered in the network from the input.}
  \label{fig:skiptrim}
\end{figure*}

Deep neural networks often rely on skip connections to address vanishing gradient problems during training~\cite{szegedy2017inception}. Skip connections provide a direct path for error propagation, allowing gradients to flow without vanishing~\cite{hochreiter1998vanishing}. Additionally, they prevent saturation issues in deep neural networks, making them more effective. Similarly, deep learning-based basecallers~\cite{wick2019performance,neumann2022rodan,konishi2021halcyon,xu2021fast,lou2020helix,perevsini2021nanopore,pages2022comprehensive,ambernas_zhang2021automated} use skip connections to mitigate the vanishing gradient and saturation problems. 
\gss{However, adding skip connections introduces the following three issues for hardware acceleration. First, skip connections increases the data-lifetime. The layers whose activations are reused in subsequent layers must wait for this activation reuse (or buffer the activations in memory) before accepting new input and continuing to compute.  This leads to high resource and storage requirements due to data duplication. Second, they introduce irregularity in neural network architecture as these connections span across non-adjacent layers.  Third, skip connections require additional computation to adjust the channel size to match the channel size at the non-consecutive layer's input. Thus, increasing model parameters and model size. Therefore, networks without skip connections have more regular topologies that translate better to hardware acceleration.

To address these issues, we propose \strim, a first skip connection remover for basecallers. 
\strim gradually removes skip connections using knowledge distillation (KD)~\cite{bucilua2006model,hinton2015distilling}, where a pretrained larger model (teacher) guides a smaller model (student) to maintain performance without skip connections.
As shown in \fgb{Additional file 1:} Figure~\ref{fig:skiptrim}, \strim starts with a pretrained over-parameterized model as the teacher, which is not updated during the training of the student network.  We use our final \nas model as the student network. We achieve skip removal by letting the teacher teach the student to perform well on basecalling.  At the start of every training epoch, \strim removes a skip connection from a block, starting from the input side, while performing KD. This is done until all skip connections are removed from the student network.  \strim gets the best of both worlds: a highly accurate and topologically regular neural network without skip connections.

During the \strim, we perform a forward pass of both the student and the teacher model, while we perform a backward pass only for the student model to update its weights. The loss to update the student network's weight during the backward pass ($\mathcal{L}_{\strim}$) is calculated with Equation~\ref{eqn:loss_skip_final}, where we use a weighing of the actual student loss ($\mathcal{L}_{S}$) and distillation loss ($\mathcal{L}_{D}$) using an alpha ($\alpha$) hyper-parameter.  The student and the teacher model compute probabilities $f_{T}$ and $f_{S}$ for output labels (i.e., nucleotides A, C, G, T) in the forward pass, respectively.  We use cross entropy ($\mathcal{L}_{CR}$) in the probability distributions to calculate the distillation loss ($\mathcal{L}_{D}$) as in Equation~\ref{eqn:loss_skip_distill}. The temperature ($\tau$) variable is used for \emph{softening} the probability distributions, i.e., it  controls the weight of knowledge from the teacher network for a student network to absorb. As we raise the $\tau$, the resulting soft label probability distribution becomes richer in information. 
}
\begin{align}
\mathcal{L}_{\strim}=\alpha\mathcal{L}_{S}-(1-\alpha)\mathcal{L}_{D} \label{eqn:loss_skip_final}\\
where \mathcal{L}_{D}= \mathcal{L}_{CR}(f_{T}/\tau,f_{S}/\tau)\label{eqn:loss_skip_distill}
\end{align} 
 
\subsection{Sensitivity to skip connection} 
\label{suppsec:skipconnection}
\begin{figure}[h]
  \centering
\includegraphics[width=0.65\linewidth,trim={0.2cm 0.25cm 0.2cm 0cm},clip]{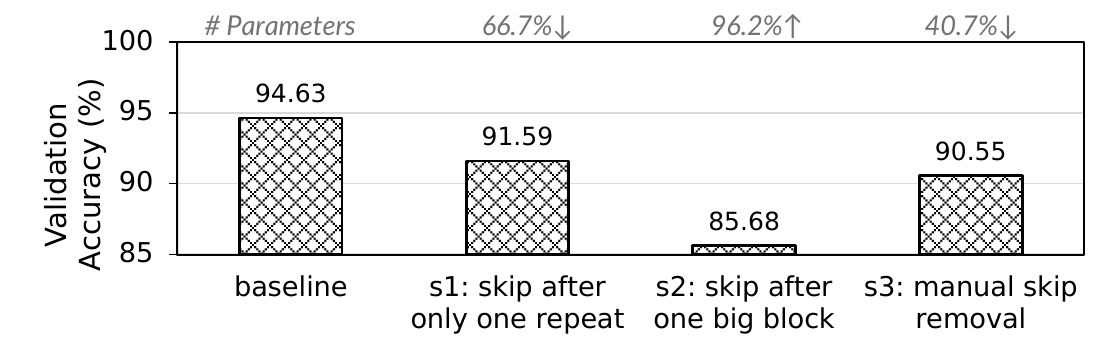}
  \caption{Basecaller sensitivity to skip connections.}
  \label{supfig:skip_sensitivity}
\end{figure}

Many state-of-the-art deep learning-based basecallers~\cite{wick2019performance,neumann2022rodan,konishi2021halcyon,xu2021fast,lou2020helix,perevsini2021nanopore,pages2022comprehensive,ambernas_zhang2021automated} incorporate skip connections to improve their basecalling accuracy. \fgb{Additional file 1:} Figure~\ref{supfig:skip_sensitivity} shows the accuracy of \bon using two different configurations of skip connections (\texttt{s1} and \texttt{s2}) and one configuration without any skip connections (\texttt{s3}) and compares it to the baseline \bon architecture. In \texttt{s1} configuration, we reduce the number of repeats in each block to one, while in \texttt{s2} configuration, we use only one block with maximum channel size, maximum kernel size, and the maximum number of repeats.

For \texttt{s3} configuration, we manually remove all the skip connections from each block in \bon. We also annotate the change in model parameters compared to the baseline model. \bon architecture comprises several blocks, each consisting of a time channel separable convolution sub-block (referred to as repeat). We make two major observations. First, the number of sub-blocks we provide skip connection plays an important role. In \texttt{s1} configuration, we observe that by using only one repeat, we reduce the accuracy by 2.84\% with 66.7\% lower model parameters, while by merging all the blocks into one big block in \texttt{s2} configuration, we observe 8.75\% lower accuracy with 96.2\% higher model parameters. Second, manually removing all the skip connections in \texttt{s3} configuration leads to 40.7\% lower model parameters at the expense of a 3.88\% loss in accuracy. This performance degradation is because, during neural network training, these connections provide a direct path for propagating the error through the layers and dealing with the vanishing gradient problem, allowing deep networks to learn properly and converge during training. Therefore, manual removal of skip connections can lead to lower basecalling performance. We conclude that skip connections are critical for basecalling accuracy.

\subsection{Hyper-parameter tuning for \strim}
\label{suppsec:Hyper_skipclip}
\begin{figure}[h]
  \centering
\includegraphics[width=0.6\linewidth,trim={0.2cm 0.25cm 0.4cm 0cm},clip]{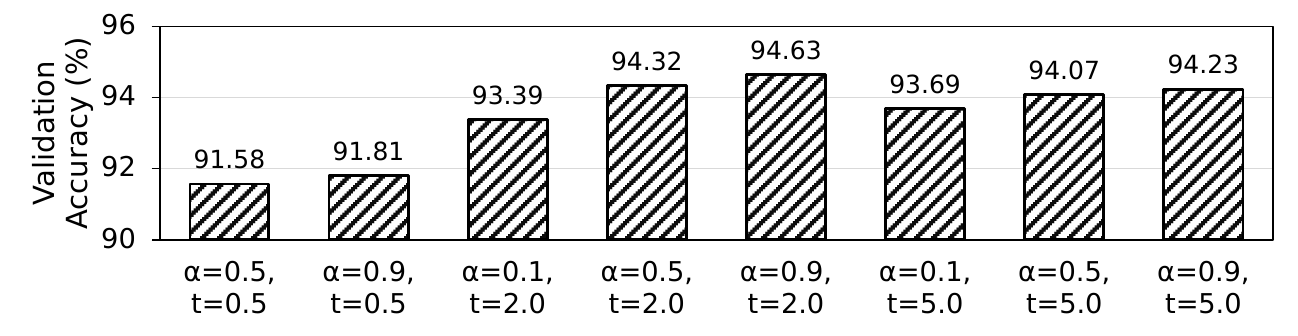}
  \caption{Sensitivity of \strim to hyper-parameters alpha ($\alpha$) and temperature ($\tau$).}
  \label{supfig:kd_hyper}
  \vspace{12pt}
\end{figure}

\Copy{IR1.10}{In \fgb{Additional file 1:} Figure~\ref{supfig:kd_hyper}, we show the effect of two critical hyper-parameters of \strim (alpha ($\alpha$) and temperature ($\tau$)) on \hgb{the validation accuracy of \bon}.}  We observe that as we raise $\alpha$ while keeping $\tau$ constant, the basecaller accuracy increases. At higher $\alpha$, \strim gives more importance to the student loss than the distillation loss during the backward pass. We use $\alpha$ =0.9 throughout our experiments.

For $\tau$, we experiment with values ranging from 0.5 to 5.0. Increasing $\tau$ provides more knowledge from the teacher network for a student network to absorb. We observe at $\tau$=2, \strim provides the highest accuracy. Further increasing $\tau$ does not provide benefits because the student network cannot absorb knowledge provided by the teacher network.

\section{\mech architecture}
\label{supsec:rubicall_arch}
\begin{figure}[h]
  \centering
  \includegraphics[width=0.5\linewidth]{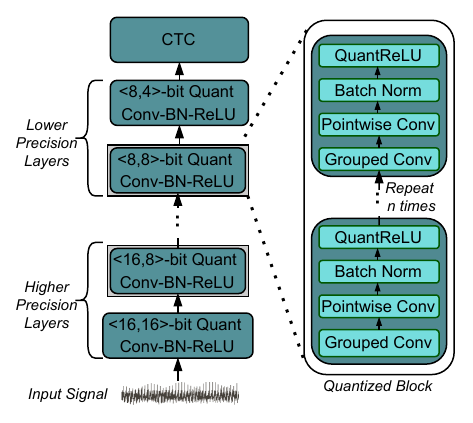}
  \caption{Overview of \mech architecture. The normalized input signal is passed through a succession of quantized convolution blocks. Each block is composed of several processing steps (convolution, batch normalization, and activation). We represent the quantization as a tuple $<$weight, activation$>$. Initial layers use a higher precision for weights and activations, while the final layers use a lower precision. The final output is passed through a connectionist temporal classification (CTC) to produce the decoded sequence of nucleotides.}
  \label{fig:rubicon}
\end{figure}

\fgb{Additional file 1:} Figure~\ref{fig:rubicon} shows the architecture of \mech. We develop \mech using \nas and \strim. The \mech architecture is composed of 28 quantized convolution blocks containing $\sim$3.3 million model parameters. Each block consists of  quantized grouped 1-dimensional convolution and quantized pointwise 1-dimensional convolution where every layer is quantized to a different domain. The convolution operation is followed by batch normalization (Batch Norm)~\cite{ioffe2015batch} and a quantized rectified linear unit (QuantReLU)~\cite{agarap2018deep} activation function. The final output is passed through a connectionist temporal classification (CTC)~\cite{graves2006connectionist} layer to produce the decoded sequence of nucleotides (A, C, G, T). CTC is used to provide the correct alignment between the input and the output sequence.  

In a learning task, $\mathcal{X}$ represents feature space with label $\mathcal{Y}$, where a machine learning model is responsible for estimating a function $f$: $\mathcal{X} \to \mathcal{Y}$.  \mech  first splits a long read in electrical-signal format (e.g., millions of signals) into multiple smaller chunks (e.g., thousands of samples per chunk) and then basecalls these chunks.  \mech uses the input signal (or squiggle) as $\mathcal{X}$ to predict nucleotides as label $\mathcal{Y}$. The CTC layer assigns a probability for all possible labels in  $\mathcal{Y}$ given an $\mathcal{X}$ at each time-step. The nucleotide with the highest probability is selected as the final output.

\section{Comparison to more accurate basecallers}
\label{suppsec:guppy_compare}
\begin{figure}[h]
  \centering
  \includegraphics[width=1\linewidth]{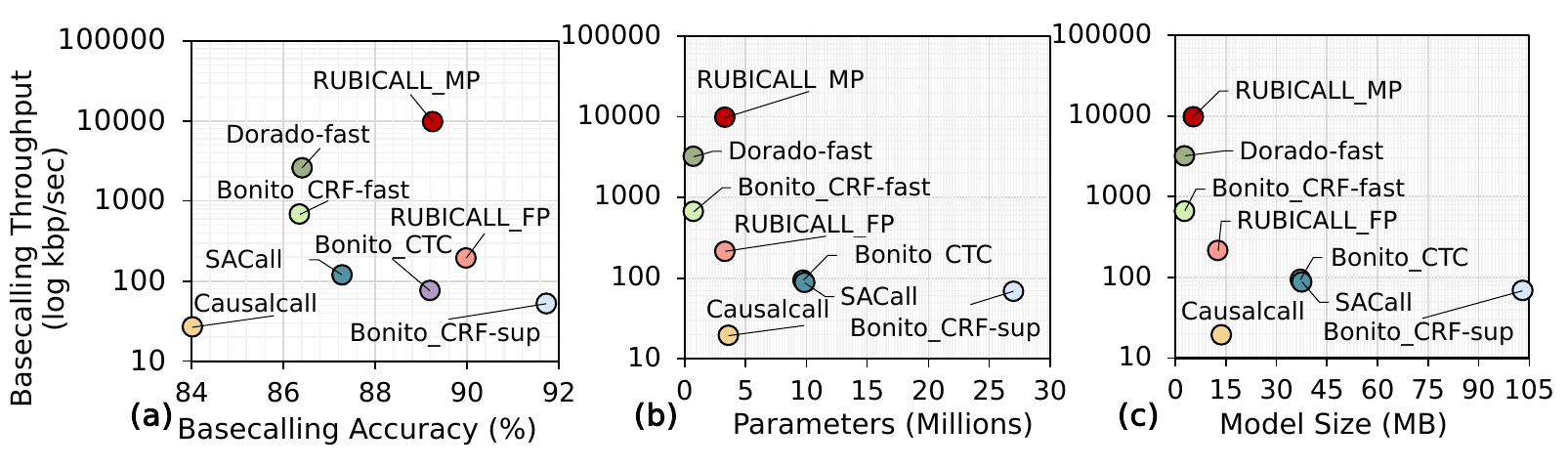}
 \caption{Comparison of \hc{average} basecalling  throughput for \mechmp with baseline basecaller in terms of:  (a) average  basecalling accuracy, (b) model parameters, and (c) model size.}
  \label{supfig:model_compare_all}
\end{figure}

Our goal is to make basecalling highly efficient and fast by building the first framework for specializing and optimizing machine learning-based basecaller. Currently, we focus on CNN-based basecallers because: (1) they are the most widely used basecallers, and  (2)  the fundamental multiply-accumulate (MAC) operation in a CNN model is amenable to hardware acceleration, unlike the operations in RNN-based basecallers. As \bon has the same backend as \mech (i.e., Quartznet~\cite{kriman2020quartznet}), we consider it as an expert-designed model. \go{\texttt{Bonito\_CRF}'s super high accuracy} (\gp) model is an RNN-based basecaller that provides more accuracy than \gpf at the expense of a \go{much larger} model. \sr{We compare} the \sr{overall} basecalling \sr{throughput} \sr{of} \mech with \sr{that of the} baseline basecaller\sr{s} in terms of basecalling accuracy, model parameters, and model size in \fgb{Additional file 1:} Figure~\ref{supfig:model_compare_all}(a), \ref{supfig:model_compare_all}(b), and \ref{supfig:model_compare_all}(c), respectively.

 \vspace{10pt}
\begin{table*}[h]
 \caption{Comparison of \mechmp with baseline basecallers in terms of model architecture, \hgb{characteristics}, precision, basecalling throughput, basecalling accuracy, parameters, and model size. For basecalling throughput and basecalling accuracy, we report average (Avg.), minimum (Min.), maximum (Max.), 25th percentile (25th \%tile), and 75th percentile (75th \%tile) values for all the basecallers.}
    \label{suptab:compare}
\centering
 \setstretch{0.9}
\renewcommand{\arraystretch}{1}
  \resizebox{1\linewidth}{!}{%
\begin{tabular}{l|l|r|l|rrrrr|rrrrr|r|r}
\hline
\textbf{Basecaller}  &  \textbf{Architecture} &  \textbf{Characteristics}  &\textbf{Precision} &\multicolumn{5}{c|}{\textbf{Basecalling Throughput (kbp/sec)}}                                            & \multicolumn{5}{c|}{\textbf{Basecalling Accuracy (\%)}}                    & \multicolumn{1}{l|}{\textbf{Parameters}} & \multicolumn{1}{l}{\textbf{\begin{tabular}[c]{@{}l@{}}Model \\ Size (MB)\end{tabular}}} \\ \hline
           &   & &      & \multicolumn{1}{r}{\textbf{Avg.}} & \multicolumn{1}{r}{\textbf{Min.}} & \multicolumn{1}{r}{\textbf{Max.}} & \multicolumn{1}{r}{\textbf{25th \%tile}} & \multicolumn{1}{c|}{\textbf{75th \%tile}} & \multicolumn{1}{r}{\textbf{Avg.}} & \multicolumn{1}{r}{\textbf{Min.}} & \multicolumn{1}{r}{\textbf{Max.}} & \multicolumn{1}{r}{\textbf{25th \%tile}} & \multicolumn{1}{r|}{\textbf{75th \%tile}} & \multicolumn{1}{r|}{--}                    & \multicolumn{1}{r}{--}                         \\
{\cc} &CNN & Low Accuracy & FP32 & 26.76 &	11.04 &	53.94 &	18.65 &	30.13                            & 84.02 &	82.70 &	86.42 &	83.57 &	86.17    & 3,589,893        & 13.69                                        \\
\sac & Transformer & Low Accuracy  &FP32& 119.71&	47.33&	346.30&	86.32&	112.58 & 87.28 &	86.44 &	91.44 &	86.97	& 89.22 & 9,854,725 & 37.59
\\
{\bon} &CNN & High Accuracy  &FP32    & 76.22&	31.68&	219.42	&54.64&	73.46                          & 89.19	& 87.99 &	93.75 &	88.62	& 91.15      & 9,738,573        & 37.15  \\
\mechfp & CNN & High Accuracy  & FP16 &194.74 &	99.8  &	394.9 &	129.72 &	240.34  & 89.25	&86.59	&93.64	&88.97 &	91.41        & 3,314,578        & 12.64    
\\
{\gp}  &RNN & Highest Accuracy & FP16    & 52.63 &	20.22	&149.2&	35.03&	55.56                                                     & \textbf{91.73} &	\textbf{90.60} &	\textbf{95.95} &	\textbf{91.43} &	\textbf{93.72 }    & 26,992,744       & 103.03                                       \\

{\gpf} &RNN & Fast Performance & FP16 & 685.13	&261.53&	1044.37&	421.52&	881.84                          & 86.36 &	82.53 &	91.39 &	86.25 &	88.51       & 730,344          & 2.79                                         \\                                
\dor & RNN & Fast Performance& FP16 & 2593.34	& 1155.06&	3927.03	&1835.99	&3351.95  & 87.16                    & 82.53 & 91.39    & 86.25       & 88.51       & 730,344          & 2.79
\\
\multirow{2}{*}{\mechmp} & \multirow{2}{*}{CNN} & High Accuracy and & Mixed- & \multirow{2}{*}{\textbf{9765.65}}&	\multirow{2}{*}{\textbf{5309.74}}&	\multirow{2}{*}{\textbf{12862.73}}&	\multirow{2}{*}{\textbf{8433.53}}&	\multirow{2}{*}{\textbf{10892.55}} & \multirow{2}{*}{89.25}	&\multirow{2}{*}{86.59}	&\multirow{2}{*}{93.64}	&\multirow{2}{*}{88.97} &	\multirow{2}{*}{91.41}       & \multirow{2}{*}{3,314,578}        & \multirow{2}{*}{5.36}\\
&  & Fast Performance & Precision &  &  &  &  &  &  & &&&& \\ \hline

\end{tabular}
}
\end{table*}

In addition to our previous observations from Figure~\ref{fig:model_compare}, we make three new observations from \fgb{Additional file 1:} Figure~\ref{supfig:model_compare_all} and \fgb{Additional file 1:} Table~\ref{suptab:compare}. \go{First, \mechmp has \rmpSpeedupSUP the performance of the highly-accurate  \gp.} \Copy{IR3.5}{\hgb{\mbox{\mechmp} is the only basecaller that provides both higher performance and accuracy \ogb{when compared to all the other evaluated basecallers}.}}
 Second, \gp uses 7.52$\times$,	36.96$\times$,	2.77$\times$, 2.74$\times$, 36.96$\times$, and	8.14$\times$ model parameters leading to a model size of 7.53$\times$,	36.93$\times$,	2.77$\times$, and	19.22$\times$ compared to \cc, \gpf, \bon, \sac, \dor and \mechmp, respectively. Third, \gp is 5.37\% more accurate than its throughout-optimized version, \gpf, which provides up to 13.02$\times$ higher basecalling performance.  We conclude that the high accuracy of a basecaller comes at a substantial cost in terms of lower throughput due to the higher number of model parameters and model size.



\begin{figure}[H]
  \centering
  \includegraphics[width=0.8\linewidth]{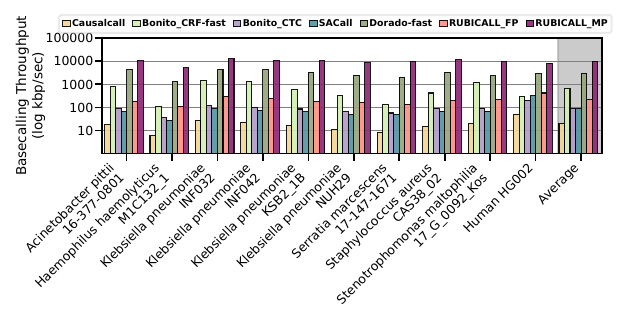}
  \vspace{-15pt}
    \caption{Performance comparison of \mech (using floating-point precision (\mechfp) and mixed-precision (\mechmp))~\sr{and five state-of-the-art basecallers on NVIDIA A40~\cite{a40}. The y-axis is on a logarithmic scale.}}
  \label{fig:model_compare_nvidia}
\end{figure}

\Copy{CC2/1}{
\section{\hgb{Evaluation on other hardware platforms}}
\label{suppsec:other_platforms}
\hgb{We also evaluate the performance of  \mech and all the other basecallers on NVIDIA A40~\mbox{\cite{a40}} GPU with 48GiB DRAM and AMD EPYC 7442~\mbox{\cite{amdEPYC}} 24-Core with 256GiB DRAM. Compared to the AMD MI210~\mbox{\cite{mi210}}, the NVIDIA A40 has a 1.65$\times$ higher peak compute performance while maintaining a 2.35$\times$ lower peak memory bandwidth. 

We make two major observations from \fgb{Additional file 1:} Figure~\mbox{\ref{fig:model_compare_nvidia}}. First, \mbox{\mechmp} on AIE consistently outperforms A40 by 502.52$\times$, 14.67$\times$, 104.14$\times$, 111.25$\times$, 45.61$\times$, and 3.19$\times$ higher performance compared to \cc, \gpf, \bon, \sac, \mechfp, and \dor, respectively. Second, for compute-bound basecallers, A40 provides 1.23$\times$, 1.18$\times$, and 1.09$\times$ higher performance than AMD MI210 (Figure~\ref{fig:perf}) for \bon, \dor, and \mechfp, respectively. \ogb{For} memory-bound basecallers,  A40 provides 1.38$\times$, 1.03$\times$, and 1.36$\times$ lower performance for \cc, \gpf, and \sac, respectively. We conclude that \mbox{\framework} provides \ogb{benefits} across \ogb{multiple hardware} platforms.}}

 %

 \Copy{IR3.9}{\hgb{
\section{Analysis of mapped reads and mapped bases}
\label{suppsec:read_lenght_analysis} \fgb{Additional file 1:} 
Table~\ref{tab:read_length_analysis} shows the average read length, the overall number of mapped reads, the number of mapped bases, and the ratio of mapped bases to the mapped reads. Our goal is to evaluate the tools in terms of the read lengths they can generate and the alignable fraction of these reads to their corresponding reference genomes. We make three key observations.
First, we find that the average read lengths are similar across different basecallers for each dataset, except \cc for the human genome. This indicates that the substantial differences in read length are unlikely to influence the ratio of mapped bases to the number of mapped reads, while the number of alignable sequences within each read and the number of mapped reads can have the main effect on such a ratio.
Second, we find that basecallers provide a similar number of mapped reads and the ratio of mapped bases to the mapped reads for each dataset, except \cc for the human genome. These similarities mainly indicate that the unalignable reads and the unalignable regions within each read are likely to be similar across basecallers, leading to similar ratios of mapped bases to mapped reads when mapping reads with similar average read lengths. 
Third, we find that \cc provides exceptions for the human genome in terms of the average read length and the mapped bases to the mapped reads ratio. This is mainly because \cc fails to basecall all raw signals for the human genome and provides a subset of basecalled reads that other basecallers generate, leading to inaccurate analysis overall.
We conclude that almost all basecallers, except \cc, generate reads with similar average read lengths and reads with similar alignable regions, although these similarities differ by certain percentages' as we discuss in Section~\ref{subsubsection:down_read_mapping}.
}}

\begin{table*}[htbp]
 \caption{Read mapping comparison of \mech with baseline basecallers in terms of mean length of individual sequencing reads in a dataset (Avg. Length), the total number of mapped reads (Mapped Reads), the total number of mapped bases (Mapped Bases), and the ratio of total number of mapped reads to mapped bases.}
     \label{tab:read_length_analysis}
\centering
 \setstretch{0.8}
\renewcommand{\arraystretch}{1}
  \resizebox{0.85\linewidth}{!}{%
\begin{tabular}{llllll}
\hline
\textbf{Dataset}                                                                                             & \textbf{Basecaller} & \textbf{Avg. Length}         & \textbf{Mapped Reads}       & \textbf{Mapped Bases}              & \textbf{\begin{tabular}[c]{@{}l@{}}\#Mapped Bases/\\ \#Mapped Reads\end{tabular}} \\\hline
\multirow{6}{*}{\begin{tabular}[c]{@{}l@{}}Acinetobacter \\ pittii 16-377-0801\end{tabular}}                 & causalcall          & \multicolumn{1}{r}{25,718.6} & \multicolumn{1}{r}{4,434}   & \multicolumn{1}{r}{114,159,528}    & \multicolumn{1}{r}{25,746.4}                                                      \\
             & \gpf            & \multicolumn{1}{r}{26,151.3} & \multicolumn{1}{r}{4,452}   & \multicolumn{1}{r}{110,907,740}    & \multicolumn{1}{r}{24,911.9}                                                      \\
             & \bon              & \multicolumn{1}{r}{24,879.1} & \multicolumn{1}{r}{4,457}   & \multicolumn{1}{r}{110,183,466}    & \multicolumn{1}{r}{24,721.4}                                                      \\
             & \sac              & \multicolumn{1}{r}{25,153.3} & \multicolumn{1}{r}{4,451}   & \multicolumn{1}{r}{111,997,940}    & \multicolumn{1}{r}{25,162.4}                                                      \\
             & \dor              & \multicolumn{1}{r}{26,151.1} & \multicolumn{1}{r}{4,452}   & \multicolumn{1}{r}{110,907,740}    & \multicolumn{1}{r}{24,911.9}                                                      \\
             & \mech            & \multicolumn{1}{r}{25,000.8} & \multicolumn{1}{r}{4,452}   & \multicolumn{1}{r}{111,405,897}    & \multicolumn{1}{r}{25,023.8}                                                      \\
             \hline
\multirow{6}{*}{\begin{tabular}[c]{@{}l@{}}Haemophilus \\ haemolyticus \\ M1C132\_1\end{tabular}}               & causalcall          & \multicolumn{1}{r}{NA}       & \multicolumn{1}{r}{NA}      & \multicolumn{1}{r}{NA}             & \multicolumn{1}{r}{NA}                                                            \\
             & \gpf            & \multicolumn{1}{r}{9,835.7}  & \multicolumn{1}{r}{6,444}   & \multicolumn{1}{r}{64,816,196}     & \multicolumn{1}{r}{10,058.4}                                                      \\
             & \bon              & \multicolumn{1}{r}{8,862.8}  & \multicolumn{1}{r}{6,201}   & \multicolumn{1}{r}{73,573,092}     & \multicolumn{1}{r}{11,864.7}                                                      \\
             & \sac              & \multicolumn{1}{r}{6,871.1}  & \multicolumn{1}{r}{4,028}   & \multicolumn{1}{r}{43,233,160}     & \multicolumn{1}{r}{10,733.2}                                                      \\
             & \dor              & \multicolumn{1}{r}{9,844.8}  & \multicolumn{1}{r}{6,444}   & \multicolumn{1}{r}{64,816,196}     & \multicolumn{1}{r}{10,058.4}                                                      \\
             & \mech            & \multicolumn{1}{r}{7,751.5}  & \multicolumn{1}{r}{6,287}   & \multicolumn{1}{r}{63,415,299}     & \multicolumn{1}{r}{10,086.7}                                                      \\ \hline
\multirow{6}{*}{\begin{tabular}[c]{@{}l@{}}Klebsiella \\ pneumoniae \\ INF032\end{tabular}}                  & causalcall          & \multicolumn{1}{r}{35,781.9} & \multicolumn{1}{r}{15,150}  & \multicolumn{1}{r}{542,123,428}    & \multicolumn{1}{r}{35,783.7}                                                      \\
             & \gpf            & \multicolumn{1}{r}{36,556.4} & \multicolumn{1}{r}{15,147}  & \multicolumn{1}{r}{533,045,454}    & \multicolumn{1}{r}{35,191.5}                                                      \\
             & \bon              & \multicolumn{1}{r}{35,189.1} & \multicolumn{1}{r}{15,152}  & \multicolumn{1}{r}{519,659,064}    & \multicolumn{1}{r}{34,296.4}                                                      \\
             & \sac              & \multicolumn{1}{r}{35,078.5} & \multicolumn{1}{r}{15,150}  & \multicolumn{1}{r}{531,456,488}    & \multicolumn{1}{r}{35,079.6}                                                      \\
             & \dor              & \multicolumn{1}{r}{36,624.7} & \multicolumn{1}{r}{15,147}  & \multicolumn{1}{r}{533,045,454}    & \multicolumn{1}{r}{35,191.5}                                                      \\
             & \mech            & \multicolumn{1}{r}{35,420.7} & \multicolumn{1}{r}{15,153}  & \multicolumn{1}{r}{536,724,002}    & \multicolumn{1}{r}{35,420.3}                                                      \\\hline
\multirow{6}{*}{\begin{tabular}[c]{@{}l@{}}Klebsiella \\ pneumoniae \\ INF042\end{tabular}}                  & causalcall          & \multicolumn{1}{r}{48,483.8} & \multicolumn{1}{r}{11,236}  & \multicolumn{1}{r}{542,123,428}    & \multicolumn{1}{r}{48,248.8}                                                      \\
             & \gpf            & \multicolumn{1}{r}{49,617.5} & \multicolumn{1}{r}{11,252}  & \multicolumn{1}{r}{533,045,454}    & \multicolumn{1}{r}{47,373.4}                                                      \\
             & \bon              & \multicolumn{1}{r}{46,198.4} & \multicolumn{1}{r}{11,273}  & \multicolumn{1}{r}{519,659,064}    & \multicolumn{1}{r}{46,097.7}                                                      \\
             & \sac              & \multicolumn{1}{r}{46,298.3} & \multicolumn{1}{r}{11,198}  & \multicolumn{1}{r}{531,456,488}    & \multicolumn{1}{r}{47,459.9}                                                      \\
             & \dor              & \multicolumn{1}{r}{49,621.4} & \multicolumn{1}{r}{11,252}  & \multicolumn{1}{r}{533,045,454}    & \multicolumn{1}{r}{47,373.4}                                                      \\
             & \mech            & \multicolumn{1}{r}{46,637.6} & \multicolumn{1}{r}{11,268}  & \multicolumn{1}{r}{536,724,002}    & \multicolumn{1}{r}{47,632.6}                                                      \\\hline
\multirow{6}{*}{\begin{tabular}[c]{@{}l@{}}Klebsiella \\ pneumoniae \\ KSB2\_1B\end{tabular}}                & causalcall          & \multicolumn{1}{r}{24,039.6} & \multicolumn{1}{r}{16,642}  & \multicolumn{1}{r}{401,041,491}    & \multicolumn{1}{r}{24,098.2}                                                      \\
             & \gpf            & \multicolumn{1}{r}{24,723.9} & \multicolumn{1}{r}{16,744}  & \multicolumn{1}{r}{384,436,100}    & \multicolumn{1}{r}{22,959.6}                                                      \\
             & \bon              & \multicolumn{1}{r}{22,918.8} & \multicolumn{1}{r}{16,803}  & \multicolumn{1}{r}{385,157,295}    & \multicolumn{1}{r}{22,921.9}                                                      \\
             & \sac              & \multicolumn{1}{r}{22,917.5} & \multicolumn{1}{r}{16,371}  & \multicolumn{1}{r}{381,266,978}    & \multicolumn{1}{r}{23,289.2}                                                      \\
             & \dor              & \multicolumn{1}{r}{24,728.0} & \multicolumn{1}{r}{16,744}  & \multicolumn{1}{r}{384,436,100}    & \multicolumn{1}{r}{22,959.6}                                                      \\
             & \mech            & \multicolumn{1}{r}{23,141.9} & \multicolumn{1}{r}{16,783}  & \multicolumn{1}{r}{388,897,351}    & \multicolumn{1}{r}{23,172.1}                                                      \\\hline
\multirow{6}{*}{\begin{tabular}[c]{@{}l@{}}Klebsiella \\ pneumoniae \\ NUH29\end{tabular}}                   & causalcall          & \multicolumn{1}{r}{16,233.7} & \multicolumn{1}{r}{14,954}  & \multicolumn{1}{r}{243,112,795}    & \multicolumn{1}{r}{16,257.4}                                                      \\
             & \gpf            & \multicolumn{1}{r}{16,435.7} & \multicolumn{1}{r}{15,056}  & \multicolumn{1}{r}{229,123,038}    & \multicolumn{1}{r}{15,218.1}                                                      \\
             & \bon              & \multicolumn{1}{r}{15,182.0} & \multicolumn{1}{r}{15,152}  & \multicolumn{1}{r}{233,135,041}    & \multicolumn{1}{r}{15,386.4}                                                      \\
             & \sac              & \multicolumn{1}{r}{15,536.5} & \multicolumn{1}{r}{15,088}  & \multicolumn{1}{r}{234,764,649}    & \multicolumn{1}{r}{15,559.7}                                                      \\
             & \dor              & \multicolumn{1}{r}{16,419.0} & \multicolumn{1}{r}{15,056}  & \multicolumn{1}{r}{229,123,038}    & \multicolumn{1}{r}{15,218.1}                                                      \\
             & \mech            & \multicolumn{1}{r}{15,300.8} & \multicolumn{1}{r}{15,113}  & \multicolumn{1}{r}{231,523,267}    & \multicolumn{1}{r}{15,319.5}                                                      \\\hline
\multirow{6}{*}{\begin{tabular}[c]{@{}l@{}}Serratia \\ marcescens \\ 17-147-1671\end{tabular}}               & causalcall          & \multicolumn{1}{r}{8,198.0}  & \multicolumn{1}{r}{12,729}  & \multicolumn{1}{r}{104,864,058}    & \multicolumn{1}{r}{8,238.2}                                                       \\
             & \gpf            & \multicolumn{1}{r}{8,456.2}  & \multicolumn{1}{r}{16,667}  & \multicolumn{1}{r}{133,916,776}    & \multicolumn{1}{r}{8,034.8}                                                       \\
             & \bon              & \multicolumn{1}{r}{8,024.8}  & \multicolumn{1}{r}{16,715}  & \multicolumn{1}{r}{133,754,055}    & \multicolumn{1}{r}{8,002.0}                                                       \\
             & \sac              & \multicolumn{1}{r}{8,167.1}  & \multicolumn{1}{r}{16,665}  & \multicolumn{1}{r}{136,289,479}    & \multicolumn{1}{r}{8,178.2}                                                       \\
             & \dor              & \multicolumn{1}{r}{8,465.1}  & \multicolumn{1}{r}{16,667}  & \multicolumn{1}{r}{133,916,776}    & \multicolumn{1}{r}{8,034.8}                                                       \\
             & \mech            & \multicolumn{1}{r}{8,076.5}  & \multicolumn{1}{r}{16,696}  & \multicolumn{1}{r}{134,916,360}    & \multicolumn{1}{r}{8,080.8}                                                       \\\hline
\multirow{6}{*}{\begin{tabular}[c]{@{}l@{}}Staphylococcus \\ aureus \\ CAS38\_02\end{tabular}}               & causalcall          & \multicolumn{1}{r}{21,425.2} & \multicolumn{1}{r}{11,038}  & \multicolumn{1}{r}{236,529,129}    & \multicolumn{1}{r}{21,428.6}                                                      \\
             & \gpf            & \multicolumn{1}{r}{21,932.6} & \multicolumn{1}{r}{11,047}  & \multicolumn{1}{r}{237,020,597}    & \multicolumn{1}{r}{21,455.7}                                                      \\
             & \bon              & \multicolumn{1}{r}{21,455.7} & \multicolumn{1}{r}{11,047}  & \multicolumn{1}{r}{232,091,657}    & \multicolumn{1}{r}{21,009.5}                                                      \\
             & \sac              & \multicolumn{1}{r}{21,372.6} & \multicolumn{1}{r}{11,047}  & \multicolumn{1}{r}{236,103,648}    & \multicolumn{1}{r}{21,372.6}                                                      \\
             & \dor              & \multicolumn{1}{r}{21,930.8} & \multicolumn{1}{r}{11,047}  & \multicolumn{1}{r}{237,020,597}    & \multicolumn{1}{r}{21,455.7}                                                      \\
             & \mech            & \multicolumn{1}{r}{21,501.0} & \multicolumn{1}{r}{11,047}  & \multicolumn{1}{r}{237,521,476}    & \multicolumn{1}{r}{21,501.0}                                                      \\\hline
\multirow{6}{*}{\begin{tabular}[c]{@{}l@{}}Stenotrophomonas \\ maltophilia \\ 17\_G\_0092\_Kos\end{tabular}} & causalcall          & \multicolumn{1}{r}{31,415.3} & \multicolumn{1}{r}{15,946}  & \multicolumn{1}{r}{501,018,868}    & \multicolumn{1}{r}{31,419.7}                                                      \\
             & \gpf            & \multicolumn{1}{r}{31,736.6} & \multicolumn{1}{r}{15,959}  & \multicolumn{1}{r}{470,408,299}    & \multicolumn{1}{r}{29,476.1}                                                      \\
             & \bon              & \multicolumn{1}{r}{29,453.8} & \multicolumn{1}{r}{15,997}  & \multicolumn{1}{r}{474,913,094}    & \multicolumn{1}{r}{29,687.6}                                                      \\
             & \sac              & \multicolumn{1}{r}{30,095.7} & \multicolumn{1}{r}{15,985}  & \multicolumn{1}{r}{481,168,698}    & \multicolumn{1}{r}{30,101.3}                                                      \\
             & \dor              & \multicolumn{1}{r}{31,727.6} & \multicolumn{1}{r}{15,959}  & \multicolumn{1}{r}{470,408,299}    & \multicolumn{1}{r}{29,476.1}                                                      \\
             & \mech            & \multicolumn{1}{r}{29,676.7} & \multicolumn{1}{r}{15,980}  & \multicolumn{1}{r}{474,401,853}    & \multicolumn{1}{r}{29,687.2}                                                      \\\hline
\multirow{6}{*}{\begin{tabular}[c]{@{}l@{}}Human \\ HG002\end{tabular}}                                      & causalcall          & \multicolumn{1}{r}{11,201.7} & \multicolumn{1}{r}{163,984} & \multicolumn{1}{r}{2,612,902,733}  & \multicolumn{1}{r}{15,933.9}                                                      \\
             & \gpf            & \multicolumn{1}{r}{37,755.9} & \multicolumn{1}{r}{238,205} & \multicolumn{1}{r}{10,627,000,000} & \multicolumn{1}{r}{44,612.8}                                                      \\
             & \bon              & \multicolumn{1}{r}{35,831.5} & \multicolumn{1}{r}{243,686} & \multicolumn{1}{r}{10,212,000,000} & \multicolumn{1}{r}{41,906.4}                                                      \\
             & \sac              & \multicolumn{1}{r}{32,815.7} & \multicolumn{1}{r}{203,228} & \multicolumn{1}{r}{9,080,197,989}  & \multicolumn{1}{r}{44,679.9}                                                      \\
             & \dor              & \multicolumn{1}{r}{37,991.9} & \multicolumn{1}{r}{238,474} & \multicolumn{1}{r}{10,627,000,000} & \multicolumn{1}{r}{44,562.5}                                                      \\
             & \mech            & \multicolumn{1}{r}{37,141.2} & \multicolumn{1}{r}{245,373} & \multicolumn{1}{r}{10,591,000,000} & \multicolumn{1}{r}{43,162.9}                                                      \\ \hline

\end{tabular}
}
\end{table*}

 \Copy{IR3.6}{\ogb{
\section{K-mer counting analysis}
\label{suppsec:kmer_analysis}

We analyze the occurrence of k-mer (i.e., substrings of length k) in a given sequence of basecalled reads and their assemblies in \fgb{Additional file 1:} Figure~\ref{fig:kmer_reads} and \fgb{Additional file 1:} Figure~\ref{fig:kmer_assembly}, respectively. We use BBMap~\cite{bbmap} to collect the number of unique k-mers and the frequency of each unique k-mer in a given sequence. During our analysis, we vary the value of k from 15 to 31. Based on our empirical analysis, we set the k value for our evaluated bacterial species to 15, where we observe distinct peaks of unique k-mers. We do not perform k-mer frequency analysis for the human genome due to the low coverage of the human genome in our experiments. 
We make the following two observations from \fgb{Additional file 1:} Figure~\ref{fig:kmer_reads} and \fgb{Additional file 1:} Figure~\ref{fig:kmer_assembly}. First, \mech has distinct peaks for all the evaluated species, often matching the k-mer composition generated from \bon. Second, \gpf and \dor generate similar k-mer compositions as they both have the same neural network architecture. \\

\begin{figure}[h]
  \centering
  \includegraphics[width=1\linewidth]{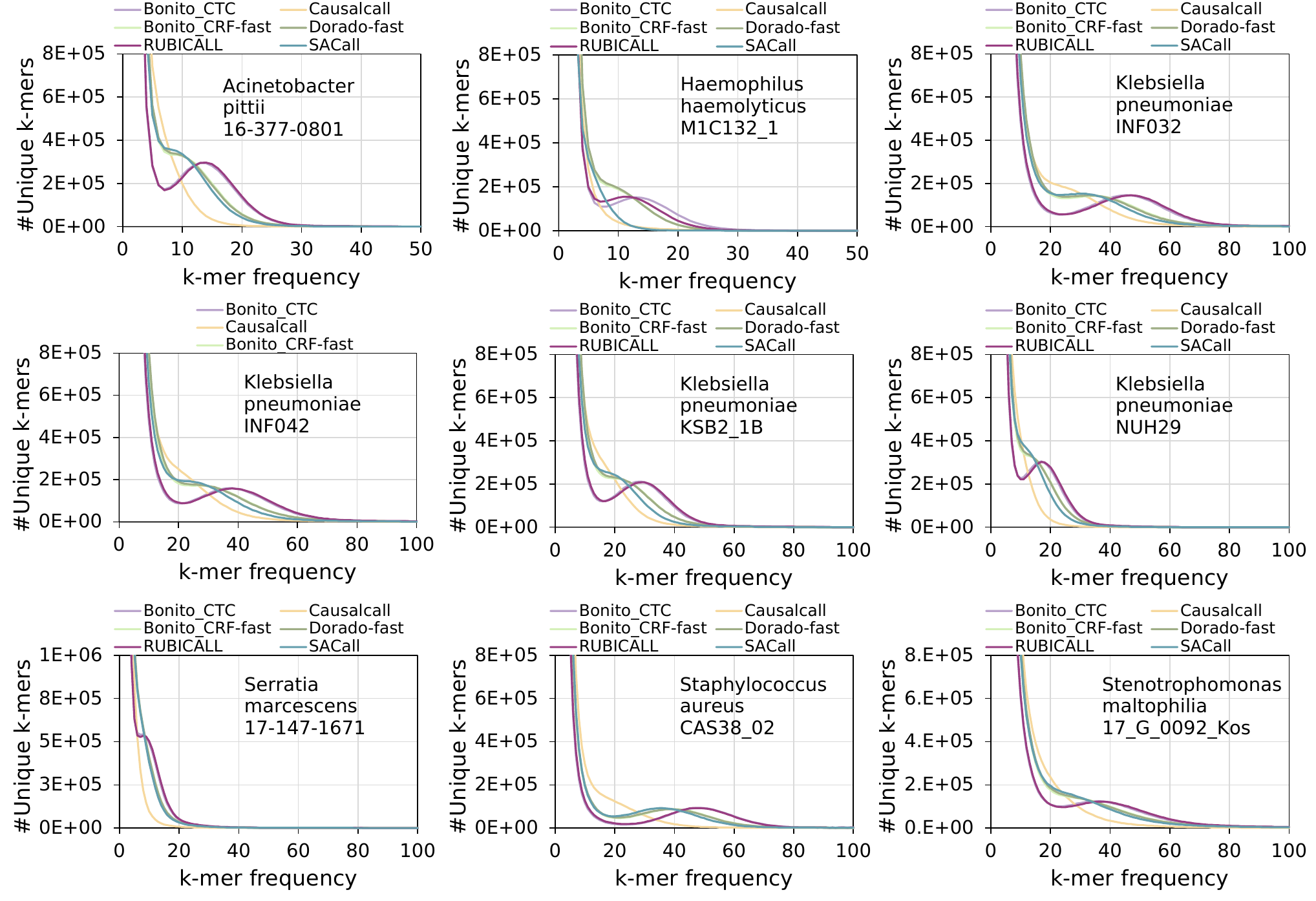}
    \caption{K-mer frequency analysis of generated reads from \mech and all the other evaluated basecallers.}
  \label{fig:kmer_reads}
\end{figure}

\begin{figure}[H]
  \centering
  \includegraphics[width=1\linewidth]{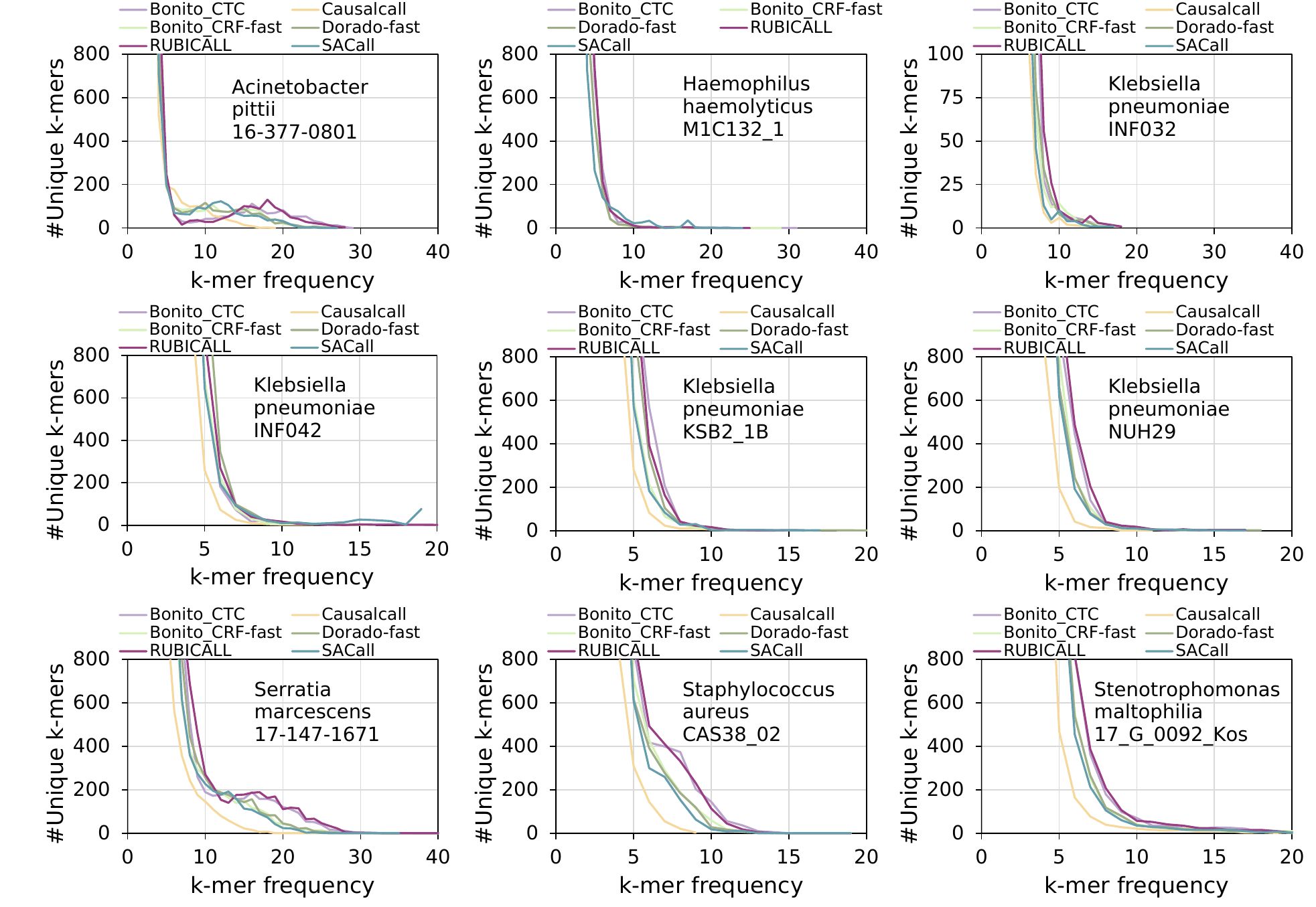}
    \caption{K-mer frequency analysis of generated assemblies of reads from \mech and all the other evaluated basecallers.}
  \label{fig:kmer_assembly}
\end{figure}

\fgb{Additional file 1:} Table~\ref{tab:kmer_compare} presents an analysis of k-mer frequencies in the raw reads and the corresponding assemblies. We include common sequences and read-to-assembly ratios to provide a comprehensive view of the similarities and disparities in sequence representation, aiding in assessing data quality and the performance of the assembly algorithms. We observe that the k-mers identified as over-represented in the assemblies are mainly observed as over-represented k-mers in read sets for most basecallers. These over-represented k-mers are likely to appear due to the particular repetitive regions of each genome, making k-mers appear a larger amount of times for these regions. Therefore, there is potentially no additional insertion or depletion of these k-mers during the assembly process.
}
}
\begin{table*}[h]
 \caption{Comparison of under and over-represented sequences (k-mers) in reads and assemblies for all the evaluated basecallers. For both under and over-represented sequences, we show common sequences (\texttt{Common}) and the ratio of k-mer frequencies between reads and assemblies (\texttt{Ratio}).}
     \label{tab:kmer_compare}
\centering
 \setstretch{0.8}
\renewcommand{\arraystretch}{1}
  \resizebox{0.85\linewidth}{!}{%
\begin{tabular}{ll|ccrc|cccr}
\hline
&                     & \multicolumn{4}{c|}{\textbf{Under-Represented}}                                                                                                                 & \multicolumn{4}{c}{\textbf{Over-Represented}}                                                                                                                  \\ \hline
\textbf{Dataset}                                                                                               & \textbf{Basecaller} & \multicolumn{1}{c}{\textbf{Read}}              & \multicolumn{1}{c}{\textbf{Assembly}} & \multicolumn{1}{c}{\textbf{Common}} & \multicolumn{1}{l|}{\textbf{Ratio}} & \multicolumn{1}{c}{\textbf{Read}}             & \multicolumn{1}{l}{\textbf{Assembly}} & \multicolumn{1}{l}{\textbf{Common}} & \multicolumn{1}{l}{\textbf{Ratio}} \\\hline
\multirow{6}{*}{\begin{tabular}[c]{@{}l@{}}Acinetobacter \\ pittii \\ 16-377-0801\end{tabular}}              & \cc          & 73,096,681                                     & 3,768,504                             & 3,221,668                           & 0.855                                        & 6,263                                         & 187                                   & 187                                 & 1.000                                        \\
         & \gpf           & 55,359,526                                     & 3,747,817                             & 2,367,806                           & 0.632                                        & 17,983                                        & 360                                   & 360                                 & 1.000                                        \\
         & \bon              & 44,790,782                                     & 3,593,419                             & 1,814,762                           & 0.505                                        & 29,097                                        & 296                                   & 296                                 & 1.000                                        \\
         & \sac              & 55,660,535                                     & 3,625,236                             & 2,381,232                           & 0.657                                        & 15,534                                        & 368                                   & 368                                 & 1.000                                        \\
         & \dor              & 55,775,603                                     & 3,760,029                             & 2,404,108                           & 0.639                                        & 18,137                                        & 425                                   & 425                                 & 1.000                                        \\
         & \mech            & 44,085,891                                     & 3,609,296                             & 1,793,772                           & 0.497                                        & 30,316                                        & 430                                   & 430                                 & 1.000                                        \\\hline
\multirow{6}{*}{\begin{tabular}[c]{@{}l@{}}Haemophilus\\ haemolyticus\\ M1C132\_1\end{tabular}}              & \cc          & 31,021,572                                     & \multicolumn{1}{c}{NA}                & \multicolumn{1}{c}{NA}              & \multicolumn{1}{c|}{NA}                       & 1,552                                         & \multicolumn{1}{c}{NA}                & \multicolumn{1}{c}{NA}              & \multicolumn{1}{c}{NA}                       \\
         & \gpf           & 42,355,232                                     & 2,077,823                             & 2,076,203                           & 0.999                                        & 2,865                                         & 54                                    & 53                                  & 0.981                                        \\
         & \bon              & 35,847,257                                     & 1,919,667                             & 700,997                             & 0.365                                        & 33,713                                        & 94                                    & 94                                  & 1.000                                        \\
         & \sac              & 36,998,888                                     & 2,000,357                             & 1,998,679                           & 0.999                                        & 22,517                                        & 98                                    & 61                                  & 0.622                                        \\
         & \dor              & 41,939,332                                     & 2,074,317                             & 2,072,740                           & 0.999                                        & 4,714                                         & 37                                    & 37                                  & 1.000                                        \\
         & \mech            & 33,917,316                                     & 1,929,302                             & 1,127,668                           & 0.584                                        & 23,221                                        & 100                                   & 100                                 & 1.000                                        \\\hline
\multirow{6}{*}{\begin{tabular}[c]{@{}l@{}}Klebsiella\\ pneumoniae\\ INF032\end{tabular}}                    & \cc          & 197,327,230                                    & 4,850,481                             & 3,833,326                           & 0.790                                        & 15,891                                        & 6                                     & 6                                   & 1.000                                        \\
         & \gpf           & 169,124,267                                    & 4,914,464                             & 3,353,593                           & 0.682                                        & 35,175                                        & 22                                    & 22                                  & 1.000                                        \\
         & \bon              & 155,835,445                                    & 4,758,242                             & 2,718,894                           & 0.571                                        & 53,149                                        & 20                                    & 20                                  & 1.000                                        \\
         & \sac              & 176,768,581                                    & 4,747,776                             & 3,366,011                           & 0.709                                        & 28,038                                        & 12                                    & 12                                  & 1.000                                        \\
         & \dor              & 167,899,392                                    & 4,916,810                             & 3,351,520                           & 0.682                                        & 35,059                                        & 24                                    & 24                                  & 1.000                                        \\
         & \mech            & 150,348,189                                    & 4,776,357                             & 2,689,193                           & 0.563                                        & 62,356                                        & 26                                    & 26                                  & 1.000                                        \\\hline
\multirow{6}{*}{\begin{tabular}[c]{@{}l@{}}Klebsiella \\ pneumoniae \\ INF042\end{tabular}}                  & \cc          & 211,565,073                                    & 5,171,081                             & 4,222,032                           & 0.816                                        & 22,204                                        & 3                                     & 3                                   & 1.000                                        \\
         & \gpf           & 178,074,237                                    & 5,212,872                             & 3,532,743                           & 0.678                                        & 61,454                                        & 29                                    & 29                                  & 1.000                                        \\
         & \bon              & 162,568,221                                    & 4,964,190                             & 2,843,401                           & 0.573                                        & 81,948                                        & 36                                    & 36                                  & 1.000                                        \\
         & \sac              & 186,186,165                                    & 5,024,228                             & 3,609,853                           & 0.718                                        & 40,440                                        & 41                                    & 41                                  & 1.000                                        \\
         & \dor              & 174,755,139                                    & 5,508,965                             & 3,808,628                           & 0.691                                        & 63,644                                        & 23                                    & 23                                  & 1.000                                        \\
         & \mech            & 158,433,298                                    & 4,989,328                             & 2,818,523                           & 0.565                                        & 92,045                                        & 37                                    & 37                                  & 1.000                                        \\\hline
\multirow{6}{*}{\begin{tabular}[c]{@{}l@{}}Klebsiella \\ pneumoniae \\ KSB2\_1B\end{tabular}}                & \cc          & 180,267,220                                    & 5,064,568                             & 4,648,903                           & 0.918                                        & 8,844                                         & 6                                     & 6                                   & 1.000                                        \\
         & \gpf           & 152,878,553                                    & 5,122,808                             & 4,044,261                           & 0.789                                        & 23,755                                        & 16                                    & 16                                  & 1.000                                        \\
         & \bon              & 137,461,268                                    & 4,859,560                             & 3,174,990                           & 0.653                                        & 27,211                                        & 14                                    & 14                                  & 1.000                                        \\
         & \sac              & 158,736,471                                    & 4,903,831                             & 4,117,531                           & 0.840                                        & 16,090                                        & 12                                    & 12                                  & 1.000                                        \\
         & \dor              & 150,458,414                                    & 5,265,881                             & 4,164,287                           & 0.791                                        & 24,938                                        & 19                                    & 19                                  & 1.000                                        \\
         & \mech            & 134,066,854                                    & 4,876,789                             & 3,186,025                           & 0.653                                        & 31,451                                        & 17                                    & 17                                  & 1.000                                        \\\hline
\multirow{6}{*}{\begin{tabular}[c]{@{}l@{}}Klebsiella \\ pneumoniae \\ NUH29\end{tabular}}                   & \cc          & 140,405,375                                    & 5,060,601                             & 5,004,375                           & 0.989                                        & 835                                           & 1                                     & 1                                   & 1.000                                        \\
         & \gpf           & 110,315,181                                    & 5,060,829                             & 4,696,878                           & 0.928                                        & 4,320                                         & 22                                    & 22                                  & 1.000                                        \\
         & \bon              & 97,405,757                                     & 4,775,503                             & 4,182,794                           & 0.876                                        & 5,177                                         & 20                                    & 20                                  & 1.000                                        \\
         & \sac              & 112,996,097                                    & 4,844,709                             & 4,613,301                           & 0.952                                        & 2,941                                         & 16                                    & 16                                  & 1.000                                        \\
         & \dor              & 108,585,877                                    & 5,043,253                             & 4,679,036                           & 0.928                                        & 4,483                                         & 23                                    & 23                                  & 1.000                                        \\
         & \mech            & 95,201,166                                     & 4,789,453                             & 4,136,482                           & 0.864                                        & 5,645                                         & 25                                    & 25                                  & 1.000                                        \\\hline
\multirow{6}{*}{\begin{tabular}[c]{@{}l@{}}Serratia \\ marcescens \\ 17-147-1671\end{tabular}}               & \cc          & 66,514,376                                     & 5,334,807                             & 5,193,034                           & 0.973                                        & 30,238                                        & 4                                     & 4                                   & 1.000                                        \\
         & \gpf           & 63,963,265                                     & 5,399,858                             & 4,554,600                           & 0.843                                        & 61,321                                        & 4                                     & 4                                   & 1.000                                        \\
         & \bon              & 53,413,056                                     & 5,217,101                             & 3,821,412                           & 0.732                                        & 61,275                                        & 9                                     & 9                                   & 1.000                                        \\
         & \sac              & 64,337,585                                     & 5,284,226                             & 4,534,343                           & 0.858                                        & 54,872                                        & 1                                     & 1                                   & 1.000                                        \\
         & \dor              & 63,535,166                                     & 5,451,215                             & 4,583,027                           & 0.841                                        & 62,006                                        & 4                                     & 4                                   & 1.000                                        \\
         & \mech            & 51,724,568                                     & 5,243,385                             & 3,741,711                           & 0.714                                        & 64,943                                        & 9                                     & 9                                   & 1.000                                        \\\hline
\multirow{6}{*}{\begin{tabular}[c]{@{}l@{}}Staphylococcus \\ aureus \\ CAS38\_02\end{tabular}}               & \cc          & 106,477,765                                    & 2,791,690                             & 2,784,563                           & 0.997                                        & 3,375                                         & 3                                     & 3                                   & 1.000                                        \\
         & \gpf           & 72,908,007                                     & 2,813,307                             & 2,774,962                           & 0.986                                        & 12,170                                        & 33                                    & 33                                  & 1.000                                        \\
         & \bon              & 59,047,673                                     & 2,736,120                             & 2,669,644                           & 0.976                                        & 18,475                                        & 37                                    & 37                                  & 1.000                                        \\
         & \sac              & 73,372,106                                     & 2,741,321                             & 2,710,663                           & 0.989                                        & 10,487                                        & 29                                    & 29                                  & 1.000                                        \\
         & \dor              & 73,708,021                                     & 2,824,623                             & 2,786,854                           & 0.987                                        & 11,902                                        & 37                                    & 37                                  & 1.000                                        \\
         & \mech            & 58,939,315                                     & 2,739,823                             & 2,675,209                           & 0.976                                        & 18,186                                        & 84                                    & 84                                  & 1.000                                        \\\hline
\multirow{6}{*}{\begin{tabular}[c]{@{}l@{}}Stenotrophomonas \\ maltophilia \\ 17\_G\_0092\_Kos\end{tabular}} & \cc          & 183,625,102                                    & 4,653,477                             & 4,000,323                           & 0.860                                        & 25,103                                        & 132                                   & 132                                 & 1.000                                        \\
         & \gpf           & 144,980,026                                    & 4,647,752                             & 3,083,896                           & 0.664                                        & 127,258                                       & 210                                   & 210                                 & 1.000                                        \\
         & \bon              & 127,334,549                                    & 4,383,078                             & 2,495,090                           & 0.569                                        & 170,398                                       & 285                                   & 285                                 & 1.000                                        \\
         & \sac              & 139,640,543                                    & 4,422,404                             & 3,045,239                           & 0.689                                        & 112,124                                       & 201                                   & 201                                 & 1.000                                        \\
         & \dor              & 143,087,662                                    & 4,594,794                             & 3,087,855                           & 0.672                                        & 127,653                                       & 201                                   & 201                                 & 1.000                                        \\
         & \mech            & 121,924,168                                    & 4,391,308                             & 2,430,608                           & 0.554                                        & 197,883                                       & 327                                   & 327                                 & 1.000     \\ \hline                                  
\end{tabular}
}
\end{table*}
\let\noopsort\undefined
\let\printfirst\undefined
\let\singleletter\undefined
\let\switchargs\undefined

\end{document}